\documentclass[linenumbers,twocolumn]{aa}

    \makeatletter
\def\@fnsymbol#1{\ensuremath{\ifcase#1\or \dagger\or \ddagger\or
   \mathsection\or \mathparagraph\or \|\or **\or \dagger\dagger
   \or \ddagger\ddagger \else\@ctrerr\fi}}
    \makeatother
    
\usepackage{graphicx}
\usepackage{txfonts}
\usepackage{amsmath}
\usepackage{amssymb}
\usepackage{multicol}
\usepackage{bm}
\usepackage{pdflscape}
\usepackage{booktabs}
\usepackage{listings}
\usepackage{color}
\usepackage{threeparttable}
\usepackage{multirow}
\usepackage{enumerate}
\usepackage{indentfirst}
\usepackage{stackengine}
\newcommand\xrowht[2][0]{\addstackgap[.5\dimexpr#2\relax]{\vphantom{#1}}}
\usepackage{booktabs}
\usepackage{subcaption}
\usepackage[nobiblatex]{xurl}
\usepackage{txfonts}
 \usepackage[bookmarks=false, 
     pdfnewwindow=true,
     colorlinks=true,
     linkcolor=blue,
     citecolor=blue,
     filecolor=blue,
     urlcolor=blue,
     final=true
 ]{hyperref}
 \usepackage{etoolbox}
 \usepackage{color}
\usepackage{CJK}
\defcitealias{Mehdipour2019}{M19}
\definecolor{mycolor}{RGB}{215, 75, 58}

\newcommand{\xmm}{{\emph{XMM-Newton}}}

\begin{document} 
\begin{CJK*}{UTF8}{gbsn}

\title{Warm absorber outflows in radio-loud active galactic nucleus 3C~59}

   \author{Yijun~Wang\thanks{\email{wangyijun@nju.edu.cn}}
          \inst{\ref{inst1},\ref{inst2}},
          Tao~Wang\thanks{\email{taowang@nju.edu.cn}}
          \inst{\ref{inst1},\ref{inst2}},
          Junjie~Mao
          \inst{\ref{inst3}},
          Yerong~Xu
          \inst{\ref{inst4},\ref{inst13}},
          Zhicheng~He
          \inst{\ref{inst6},\ref{inst7}},
          Zheng~Zhou
          \inst{\ref{inst8}},
          Chen~Li
          \inst{\ref{inst9},\ref{inst10}},
          Yongquan~Xue
          \inst{\ref{inst6},\ref{inst7}},
          Jiayi~Chen
          \inst{\ref{inst3}},
          Fangzheng~Shi
          \inst{\ref{inst11}},
          Missagh Mehdipour
          \inst{\ref{inst12}}
          }

   \institute{School of Astronomy and Space Science, Nanjing University, 163 Xianlin Avenue, Nanjing 210023, People's Republic of China\label{inst1}
         \and Key Laboratory of Modern Astronomy and Astrophysics, Nanjing University, Ministry of Education, 163 Xianlin Avenue, Nanjing 210023, People's Republic of China\label{inst2}
         \and Department of Astronomy, Tsinghua University, Haidian DS 100084 Beijing, People's Republic of China\label{inst3}
         \and Institute of Space Sciences (ICE, CSIC), Campus UAB, Carrer de Can Magrans s/n, 08193, Barcelona, Spain\label{inst4}
         \and Institut d'Estudis Espacials de Catalunya (IEEC), Esteve Terradas 1, RDIT Building, Of. 212 Mediterranean Technology Park (PMT), 08860, Castelldefels, Spain\label{inst13}
         \and CAS Key Laboratory for Research in Galaxies and Cosmology, Department of Astronomy, University of Science and Technology of China, Hefei 230026, China\label{inst6}
         \and School of Astronomy and Space Sciences, University of Science and Technology of China, Hefei 230026, China\label{inst7}
         \and Department of Astronomy, Xiamen University, Xiamen, Fujian 361005, People's Republic of China\label{inst8}
         \and Leiden Observatory, Leiden University, PO Box 9513, 2300 RA Leiden, The Netherlands\label{inst9}
         \and SRON Netherlands Institute for Space Research, Niels Bohrweg 4, 2333 CA Leiden, the Netherlands\label{inst10}
         \and Shanghai Astronomical Observatory, Chinese Academy of Science, No.80 Nandan Road, Shanghai, China\label{inst11}
         \and Space Telescope Science Institute, 3700 San Martin Drive, Baltimore, MD 21218, USA\label{inst12}
             }

  \abstract
  {
  Both jets and ionized outflows in active galactic nuclei (AGNs) are thought to play important roles in affecting the star formation and evolution of host galaxies, but their relationship is still unclear. 
 As a pilot study, we performed a detailed spectral analysis for a radio-loud (RL) AGN 3C~59 ($z=0.1096$) by systematically considering various factors that may affect the fitting results,
  and thereby establishing a general spectral fitting strategy for subsequent research with larger sample.
  3C~59 is one rare target for simultaneously studying jets and warm absorbers that is one type of ionized outflows.
  Based on the multi-wavelength data from near-infrared (NIR) to hard X-ray bands detected by
  Dark Energy Spectroscopic Instrument, Galaxy Evolution Explorer, and XMM-Newton,
  we used \texttt{spex} code to build
  broadband continuum models and perform photoionization modeling with \texttt{pion} code to constrain the physical parameters of warm absorbers in 3C~59.
  We found two warm absorbers with ionization parameter of 
  $\log [\xi/(\rm{erg}\ \rm{cm}\ \rm{s}^{-1})] = 2.65^{+0.10}_{-0.09}$ and $1.65\pm 0.11$, respectively,
  and their outflowing velocities are $v_{\rm out} = -528^{+163}_{-222}\ \rm{km}\ \rm{s}^{-1}$ and $-228^{+121}_{-122}\ \rm{km}\ \rm{s}^{-1}$, respectively.
  These warm absorbers are located between outer torus and narrow (emission-)line region, and their positive $v_{\rm out}$-$\xi$ relation can be explained by the radiation-pressure-driven mechanism.
 We found that the estimations of these physical properties are affected by the different spectral fitting strategies, such as the inclusion of NIR to ultra-violet (UV) data, the choice of energy range of spectrum, or the composition of the spectral energy distribution.
Based on the same fitting strategy, this work presents a comparative study of outflow driven mechanism between a RL AGN (3C 59) and a radio-quiet AGN (NGC 3227),
which suggests a similar driven mechanism of their warm absorber outflows and a negligible role of jets in this process.
   }


   \authorrunning{Y.J. Wang et al.}
   \titlerunning{Warm absorbers in radio-loud AGN 3C~59}
   \maketitle

\section{Introduction}
\label{sec:intro}

Many observational proofs have implied that active galactic nuclei (AGNs) 
play an important role in affecting star formations and evolution of their host galaxies 
through high kinetic or radiative power, so-called AGN feedback \citep[e.g.,][]{Fabian2012,King2015,Fiore2017,Matzeu2023}.
The kinetic feedback through large-scale jets or ionized outflows injects large amounts of energy and matter into host galaxies and even surrounding environments, 
which makes it a very effective feedback mechanism \citep{Boehringer1993,Fabian2006}.
However, there are still many open questions about the connection between jets and ionized outflows,
which are important for our understanding on AGN feedback.
Jets are believed to be magnetically powered \citep{Blandford1977,Blandford1982}, 
while the origins and launching mechanisms of the ionized outflows are more unclear.

Warm absorber is one type of ionized outflows,
which has ionization parameter ($\xi$) ranging from $\sim 10^{-1}$ to $\sim 10^3$ $\rm{erg}\ \rm{cm}\ \rm{s}^{-1}$,
and outflowing velocity ranging from $\sim -100$ to $\sim -2000$ km s$^{-1}$ \citep[e.g.,][]{Kaastra2000,Laha2021}.
This type of outflow is usually located between the accretion disk and narrow (emission-)line region (NLR) \citep[$\sim 0.001$ pc to $\sim 1$ kpc; e.g.,][]{Blustin2005,Krongold2007,Kaastra2012,Silva2016,Mehdipour2018,Laha2021}.
Warm absorbers are usually thought to be driven by various mechanisms, such as radiation pressure \citep[e.g.,][]{Proga2004}, 
magnetic forces \citep[e.g.,][]{Blandford1982,Konigl1994,Fukumura2010}, 
or thermal pressure \citep[e.g.,][]{Begelman1983,Krolik1995,Mizumoto2019}.
Among these mechanisms, the radiation-pressure-driven mechanism may play a crucial role because warm absorbers are mainly dusty winds \citep[e.g.,][]{Yamada2024}.
Recent studies have found that nearly 50\% of nearby radio-quiet Seyfert galaxies are detected with warm absorbers \citep[e.g.,][]{Reynolds1997,Tombesi2013,Laha2014,Laha2021}, 
while warm absorbers are only found in a few radio-loud AGNs \citep{Reeves2009,Reeves2010,Torresi2010,Torresi2012,Gesu2016,Mehdipour2019,Kayanoki2024}.
Until now, a direct physical relationship between jets and warm absorbers has not been well established.
\cite{Mehdipour2019} (hereafter \citetalias{Mehdipour2019}) argued that changes of magnetic-field configuration from toroidal to poloidal may power either warm absorbers or jets, and \cite{Torresi2012} found that powerful jets may favor the escape of more massive warm absorbers.
In addition, many works have found that kpc-scale [OIII]$\lambda$5007 ionized outflows 
are driven by the expanding radio jets, 
which may be connected with the interaction between jets and ambient circumnuclear medium (ACM) \citep[e.g.,][]{Tadhunter2001,Holt2006,Holt2008,Singha2023,Peralta2023}.
However, whether jets can drive warm absorbers still remain unclear.
Therefore, more studies about warm absorbers in radio-loud AGNs are required to help us understand the underlying connection between jets and warm absorbers, and further investigate the interaction between jets and ACM at smaller scale.

The correlation between ionization parameter and outflowing velocity of warm absorbers
characterizes not only the driving mechanisms of outflows \citep[e.g.,][]{Fukumura2010,Tombesi2013}, 
but also the density profile of ACM in the center of galaxies \citep{Wang2022}.
The latter is an important tracer of the accretion history and accretion physics of 
central supermassive black holes (SMBHs) \citep[e.g.,][]{Bondi1952,Narayan1994,Frank2002,Yuan2014}.
Until now, the density profile of ACM has been estimated for only several 
nearby galaxies \citep{Wong2011,Russell2015,Miller2015,Gillessen2019},
a few tidal-disruption-event-hosting galaxies \citep{Alexander2016,Eftekhari2018,Alexander2020,Anderson2020},
and several radio-quiet AGNs \citep{Wang2022}, 
while there is a lack of studies on ACM density profile in radio-loud AGNs.
Exploring the properties of warm absorbers in radio-loud AGNs may help us investigate the density profile of their ACM, and further enhance our understanding on the SMBH accretion physics in radio-loud AGNs.

3C~59 \citep[$z=0.1096$, ][]{Eracleous2004}, a Fanaroff Riley Class II (FRII) radio-loud AGN with radio lobes extending to at least $150$ kpc \citep[][hereafter Paper I]{WangYJ2025}.
It is one of three radio-loud AGNs that shows multi-phase warm absorbers (\citetalias{Mehdipour2019}),
which is necessary for studying the evolution of outflows and density profile of ACM.
3C~59 also has a wealth of photometric data and spectral data in multi-wavelength bands (see Paper I),
which is useful for constructing broadband spectral energy distribution (SED) and making spectral analysis to obtain various physical properties.
In Paper I, we obtained various physical properties for the host galaxy of 3C~59 
and separated AGN fluxes in the optical bands from the entire galaxy,
which is greatly helpful for constraining continuum models of central engine and studying physical properties of warm absorbers in this work.
These observational results make 3C~59 a rare case to simultaneously study warm absorbers and jets.
This work will focus on the physical properties of warm absorbers in 3C~59, and their (non-)connection with jet.

In addition, different works usually adopt distinct spectral fitting strategies to estimate the physical properties of warm absorbers, such as whether to incorporate optical/UV data to constrain the continuum models, the choice of spectral coverage of each instrument, the selection of different photoionization plasma codes, and the composition of SED models. \cite{Mehdipour2016} had made a systematic comparison of various photoionization plasma codes and found that the deviation across the different codes in the ionization parameter at which ionic abundances of H-like or B-like ions peak is about 10\%, while for C-like to Fe-like ions, this deviation is about 40\%. However, until now, few works have investigated the potential impact of optical/UV data inclusion, spectral range selection, and SED compositions on spectral fitting outcomes, which should be accounted for especially when comparing or combining the results in different works. Therefore, another important goal of this work is to study these potential impacts through comparing the best-fit results between different spectral fitting strategies.
In addition, we take this work as a pilot study to validate our spectral fitting strategy, while also preparing for its future application to large-sample research.

This paper is organized as follows. 
In Section \ref{sec:obsdata}, we present the observational data used in this work and data reduction for the XMM-Newton data.
In Section \ref{sec:spectralfitSED}, we introduce the details about constructing broadband SED and making spectral analysis.
In Section \ref{sec:resultsdiscussion}, we present the best-fit results of warm absorbers and discuss their physical properties, 
such as absorption features, possible driving mechanisms, potential effects caused by different spectral fitting strategies, distance to the SMBH, and possible (non-)connection with jets.
In Section \ref{sec:summary}, we give a summary about our conclusions.
In this work, we used C-statistic \citep[][hereafter C-stat]{Cash1979,Kaastra2017} to estimate the goodness of fit.
The statistical errors of parameters are given at 1$\sigma$ (68\%) confidence level.
Throughout this paper, we assume a flat cosmology with the following parameters:
$\Omega_{\rm m}=0.3$, $\Omega_{\Lambda}=0.7$, and $H_0=70\ \rm{km}\ \rm{s}^{-1}\ \rm{Mpc}^{-1}$.

\section{Observations and data processing}
\label{sec:obsdata}

\subsection{NIR-optical-UV data}
\label{sec:NIRoptUVdata}
The clean AGN fluxes at $g$, $r$, and $z$ filters were derived from 2-dimensional decompositions 
for the Dark Energy Spectroscopic Instrument (DESI) images (see Paper I).
The reference wavelengths ($\lambda_{\rm ref}$) of $g$, $r$, and $z$ filters are 4703 {\AA}, 6176 {\AA}, and 8947 {\AA}, respectively\footnote{See details in \url{http://svo2.cab.inta-csic.es/theory/fps/}}.
We refer readers to Paper I for more details about the image decomposition and here we only give a brief introduction.
In the image decomposition with \textsc{galight} code \citep{Ding2021}, 
AGN component is modeled using a point spread function (PSF) based on stars within the detection area,
while host galaxy is modeled with a smooth S$\acute{\rm e}$rsic profile.
The DESI photometric data were corrected with the Galactic extinction and the intrinsic extinction from the host galaxy of 3C~59.
For the Galactic extinction, we considered the \cite{Fitzpatrick1999} reddening law with $R_{V} = 3.1$ based on the dust map from \cite{Schlegel1998}.
For the intrinsic extinction from the host galaxy of 3C~59,
we adopted ${\rm E(B-V)} = 0.077$ mag that was estimated through the multi-wavelength photometric data 
fitting with the Code Investigating GALaxy Emission \citep[\textsc{cigale};][]{Burgarella2005,Noll2009,Boquien2019,Yang2020,Yang2022} in Paper I.
The extinction-corrected DESI AGN fluxes at $g$, $r$, and $z$ filters are $0.17 \pm 0.04$ mJy, $0.20 \pm 0.01$ mJy, and $0.44 \pm 0.03$ mJy, respectively.

We also utilized the UV and optical photometric data detected by Galaxy Evolution Explorer (GALEX) and Optical Monitor (OM) aboard {\xmm}.
The photometric data at FUV ($\lambda_{\rm ref} \sim 1535$ \AA) and NUV ($\lambda_{\rm ref} \sim 2301$ \AA) 
filters of GALEX were derived from \cite{Morrissey2007}.
After correction for the Galactic and intrinsic extinctions, the observational fluxes at FUV and NUV filters are $0.36 \pm 0.02$ mJy and $0.43 \pm 0.02$ mJy, respectively.
The OM images were processed with the Science Analysis System (SAS v20.0.0)
\texttt{omichain} pipeline according to the standard parameters.
The OM images of 3C~59 (observation ID is 0205390201) were taken at $V$, $B$, $U$, $UVW1$, and $UVM2$ filters
with reference wavelengths of 5425 \AA, 4337 \AA, 3477 \AA, 2931 \AA, and 2327 \AA, respectively.
The OM data were also corrected with the Galactic extinction and the intrinsic extinction from the host galaxy of 3C~59.
The extinction-corrected fluxes at $V$, $B$, $U$, $UVW1$, and $UVM2$ filters are 
$1.65 \pm 0.01$ mJy, $1.08 \pm 0.005$ mJy, $0.98 \pm 0.004$ mJy, $0.77 \pm 0.004$ mJy, and $0.45 \pm 0.005$ mJy, respectively.

   \begin{table*}[h!]
   \caption{Best-fit parameters of SED models and warm absorbers for 3C~59 \label{tab:bestfitparameters}}
   \centering
   \renewcommand{\arraystretch}{1.2}
   \begin{tabular}{llllccccccccccc}
   \hline\hline\xrowht[()]{6pt}
   Component & Model & Parameter & Symbol & Value \\
   \hline\xrowht[()]{5pt}
   Disk blackbody component & \texttt{dbb} & Normalization &  $A$ ($10^{25}\ \rm{m}^2$) & 1.34 (scaled) \\
    (NIR-optical-UV)  & & Temperature & $kT_{\rm BB}$ (eV) & 3.4 (fixed) \\
   \hline\xrowht[()]{5pt}
   Warm Comptonization component  & \texttt{comt} & Normalization & $A$ ($10^{56}\ \rm{ph} \rm{s}^{-1} \rm{keV}^{-1}$) & $4.01^{+1.71}_{-1.17}$ \\
   (soft X-ray excess)  & & Seed photons temperature & $kT_0$ (eV) & $3.4$ (fixed) \\
    & & Plasma temperature & $kT_1$ (eV) & $89.00^{+4.86}_{-4.71}$ \\
    & & Optical depth & $\tau$ & 30 (fixed) \\
   \hline\xrowht[()]{5pt}
   X-ray power-law component & \texttt{pow} & Normalization & $A$ ($10^{52}\ \rm{ph} \rm{s}^{-1} \rm{keV}^{-1}$) & $5.93^{+0.21}_{-0.16}$\\
           & & Photon index & $\Gamma$ & $1.63 \pm 0.02$ \\
   \hline\xrowht[()]{5pt}
   Neutral reflection component  & \texttt{refl} & Incident power-law normalization & $A$ ($10^{52}\ \rm{ph} \rm{s}^{-1} \rm{keV}^{-1}$) & $5.94$ (coupled) \\
     & & Incident power-law photon index & $\Gamma$ & 1.63 (coupled) \\
    & & Reflection scale & $s$ & $0.21\pm 0.06$ \\
   \hline\xrowht[()]{5pt}
   Galactic neutral gas & \texttt{hot} & Hydrogen column density & $N_{\rm H}$ ($10^{20}$ cm$^{-2}$) & $6.59$ (fixed) \\
           & & Electron temperature & $kT_{\rm e}$ (eV) & $0.001$ (fixed) \\
   \hline\xrowht[()]{5pt}
   Highly-ionized warm absorber (WA$_{\rm H}$) & \texttt{pion} & Hydrogen column density & $N_{\rm H}$ ($10^{22}$ cm$^{-2}$) & $0.69^{+0.31}_{-0.17}$ \\
           & & $\log$ of ionization parameter & $\log\ [\xi\ (\rm{erg}\ \rm{cm}\ \rm{s}^{-1})]$ & $2.65^{+0.10}_{-0.09}$ \\
           & & Turbulent velocity & $\sigma_v$ & $175^{+93}_{-87}$ \\
           & & Outflowing velocity & $v_{\rm out}\ (\rm{km}\ \rm{s}^{-1})$ & $-528^{+163}_{-222}$ \\
   \hline\xrowht[()]{5pt} 
   Lowly-ionized warm absorber (WA$_{\rm L}$) & \texttt{pion} & Hydrogen column density & $N_{\rm H}$ ($10^{22}$ cm$^{-2}$) & $0.31^{+0.04}_{-0.03}$ \\
           & & $\log$ of ionization parameter & $\log\ [\xi\ (\rm{erg}\ \rm{cm}\ \rm{s}^{-1})]$ & $1.65\pm 0.11$ \\
           & & Turbulent velocity & $\sigma_v$ & $94^{+36}_{-30}$ \\
           & & Outflowing velocity & $v_{\rm out}\ (\rm{km}\ \rm{s}^{-1})$ & $-228^{+121}_{-122}$ \\
   \hline\xrowht[()]{5pt} 
   Statistical results & & Best-fit Cash statistic & $C$-stat & 2076.9 \\
    & & Expected Cash statistic & $C$-expt & 1755.9 \\
    & & Degree of freedom & DoF & 1699\\
    \hline\xrowht[()]{5pt}
    Intrinsic luminosity & & 0.3--10 keV luminosity & $L_{\rm 0.3-10\ keV}$ (erg s$^{-1}$) & $5.19\times 10^{44}$  \\
       & & 2--10 keV luminosity & $L_{\rm 2-10\ keV}$ (erg s$^{-1}$) & $2.82\times 10^{44}$ \\
       & & 1--1000 Ryd ionizing luminosity & $L_{\rm ion}$ (erg s$^{-1}$) & $1.79\times 10^{45}$ \\
       & & Bolometric luminosity & $L_{\rm bol}$ (erg s$^{-1}$) & $3.45\times 10^{45}$ \\
    \hline
   \end{tabular}
   \end{table*}

\subsection{X-ray data reduction}
The EPIC-pn and RGS data of 3C~59 (observation ID is 0205390201) were processed using the SAS v20.0.0
following the standard data analysis 
procedure\footnote{See details in \url{https://www.cosmos.esa.int/web/xmm-newton/sas-threads}}.
The cleaned event files of the EPIC-pn data were produced through the SAS task \texttt{epproc}.
The single event (``PATTERN==0'') and 10--12 keV high energy light curve (``PI>10000\&\&PI<12000'') 
were extracted from the event file, based on which flaring particle background above a count rate threshold of 0.3 counts s$^{-1}$ was filtered.
The EPIC-pn spectrum of the source was extracted from a circular region with a radius of 50$''$ centered on the source and 
the background spectrum was extracted from a nearby source-free circular region with a radius of 150$''$. 
The SAS tasks \texttt{rmfgen} and \texttt{arfgen} were used to produce response matrices files and ancillary response files, respectively.
Then we used the SAS task \texttt{rgsproc} to extract the first-order data of RGS1 and RGS2.
For each RGS, the flaring particle background larger than 0.1 counts s$^{-1}$ 
was excluded based on the light curve of CCD 9.
To improve the signal-to-noise ratio, we combined the spectra of RGS1 and RGS2 through the SAS task \texttt{rgscombine}.

\section{Spectral analysis}
\label{sec:spectralfitSED}

We make detailed spectral analysis based on the \texttt{spex} package \citep{Kaastra1996} v.3.08.00 \citep{spexversion308}.
Firstly, following \cite{Mao2022}, \cite{Kayanoki2024}, and \cite{Zhou2024}, we cross-calibrated the RGS and EPIC-pn spectra.
Through making their fluxes matched at a common energy band, we found that the RGS spectrum needs to be rescaled by a factor of 1.03 to match the EPIC-pn spectrum.
In the spectral fitting, we used the EPIC-pn spectrum in the energy range of 2--10 keV 
and RGS spectrum in the wavelength range of 6--38 {\AA},
which is consistent with the approach adopted for NGC~3783 \citep{Mao2019}, NGC~3227 \citep{Wang2022a,MaoJJ2022}, 
MR~2251-178 \citep{Mao2022}, Mrk~6 \citep{Kayanoki2024}, and RE J1034+396 \citep{Zhou2024}.
In addition, EPIC-pn spectrum was optimally binned using the \texttt{obin} command in \texttt{spex} \citep{KaastraBleeker2016}, and RGS spectrum was binned by a factor of 2 required by optimal binning.
Then the observational data from NIR to X-ray bands were fitted by the following models in \texttt{spex},
which are also summarized in Table \ref{tab:bestfitparameters}.

\begin{figure}[t!]
\center
\includegraphics[width=\linewidth, clip]{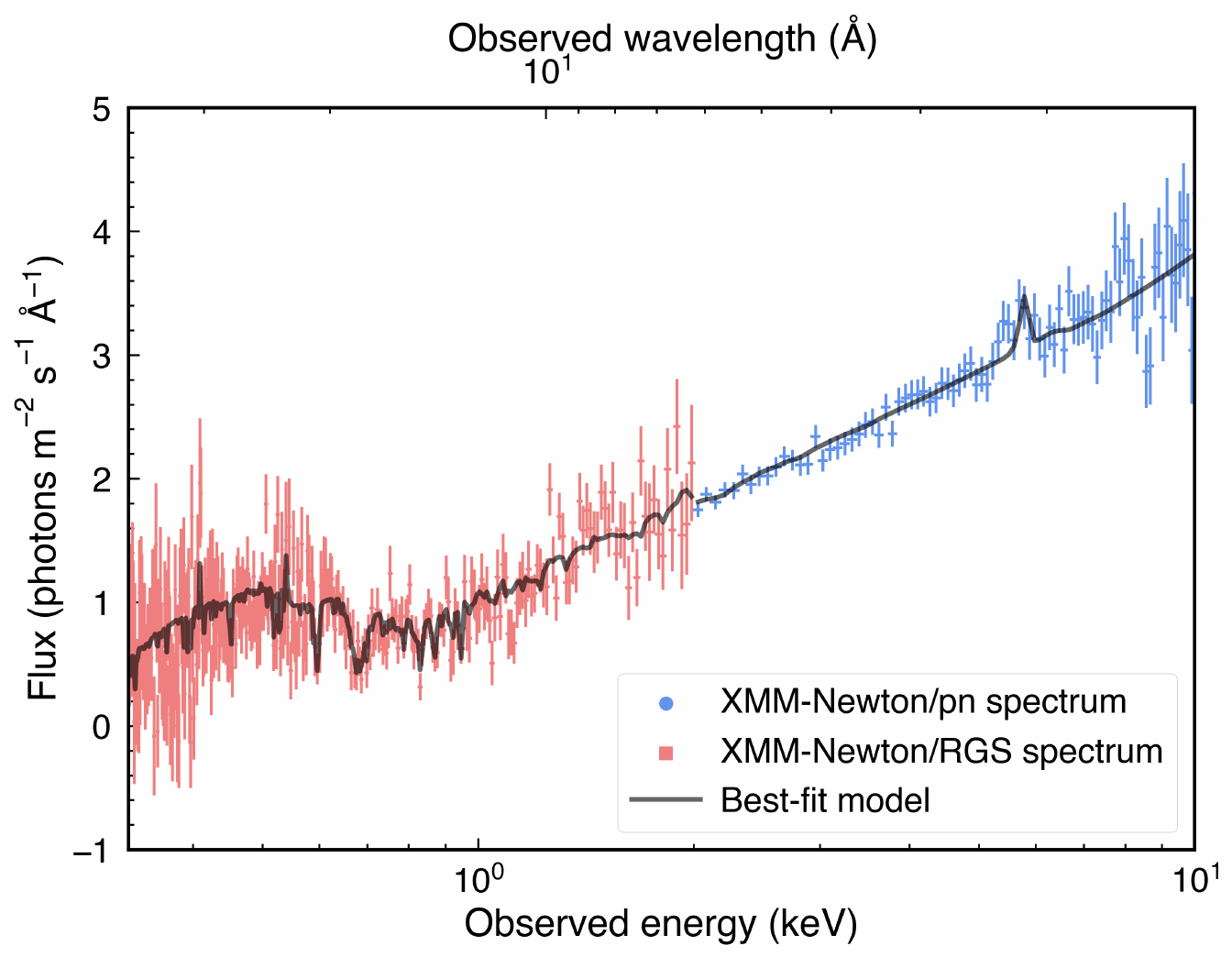}
\caption{XMM-Newton spectra and the best-fit model with \texttt{spex}.
The red points denote the RGS spectrum and
the blue points represent the EPIC-pn spectrum.
The solid black curve shows the best-fit model.
For visualization purposes, here the RGS spectrum is binned by a factor of 10.
\label{fig:dataandmodel}}
\end{figure}

\subsection{Intrinsic broadband SED}
\label{sec:continuummodel}
Similar to NGC~3227 \citep{Mehdipour2021} and MR~2251-178 \citep{Mao2022}, the intrinsic SED model of 3C~59 includes an accretion disk blackbody component, and a warm Comptonization component, an X-ray power-law component, and a neutral X-ray reflection component.

To determine the accretion disk model (\texttt{dbb} model in \texttt{spex}), NIR-optical-UV data solely from AGN are required.
Thus, AGN fluxes at DESI $g$, $r$, and $z$ filters can be directly used.
However, for GALEX and OM data, it is difficult to separate the contributions from AGN and host galaxy,
while contributions of the host galaxy cannot be 
ignored\footnote{The host galaxy of 3C~59 is an elliptical galaxy, but it is experiencing star-formation rejuvenation (see Paper I), which puts 3C~59 in a quite complex host galaxy enviroment.} (see details in Paper I).
Therefore, we only used GALEX and OM data as upper limits in this work.
In addition, the fact that only three DESI flux data points can be used does not allow us to better constrain the \texttt{dbb} model with all the parameters free. Therefore, given that the central engines of 3C~59 and 3C~382 have many similar physical properties, such as black hole masses\footnote{Based on the reverberation mapping method, the black hole mass of 3C~59 is about $10^{8.9}\ M_\odot$ \citep{Wu2004}, and it is about $10^9\ M_\odot$ for 3C~382 \citep{Fausnaugh2017}.}, bolometric luminosities, X-ray luminosities, Eddington ratios\footnote{The bolometric luminosities, 0.3--10 keV X-ray luminosities, and Eddington ratios of both 3C~59 and 3C~382 are about $4\times 10^{45}\ \rm{erg}\ \rm{s}^{-1}$, $5\times 10^{44}\ \rm{erg}\ \rm{s}^{-1}$, and 0.04, respectively \citepalias{Mehdipour2019}.}, and radio classifications\footnote{3C~59 and 3C~382 are both FRII radio-loud AGN.},
we fixed the disk temperature ($kT_{\rm BB}$) of 3C~59 to that of 3C~382, which is 3.4 eV \citep{Ursini2018}. 
Then, considering that the DESI image at $g$-filter has a deeper imaging and a higher data quality than those at $r$ and $z$ filters (see Paper I),
we directly scaled the normalization of the \texttt{dbb} model to match the extinction-corrected DESI $g$-filter AGN flux.
The resulting \texttt{dbb} model has been verified to be nearly consistent with the DESI AGN fluxes at $r$ and $z$ filters, 
and be below the fluxes of GALEX and OM data.
Here we do not utilize the spectrum (3700 {\AA} to 9000 {\AA}) detected by the Large Sky Area Multi-Object Fiber Spectroscopic Telescope (LAMOST) for 3C~59 (see Section \ref{sec:distanceWA}),
because true contributions from AGN are hard to reliably extract as we aforementioned.

Only a \texttt{dbb} model cannot reasonably match the observational data in the soft X-ray bands.
Therefore, a warm Comptonization component (\texttt{comt} model in \texttt{spex}) is introduced to explain the soft X-ray excess
that is usually thought to be produced by the up-scattering of the seed photons 
from accretion disks in a warm Comptonizing corona.
In the fit, the seed temperature ($kT_0$) was fixed to the disk temperature.
Following \cite{Mehdipour2021}, the optical depth ($\tau$) was fixed to a fiducial value of 30
in order to limit the number of free parameters and reduce the parameter degeneracy.
The normalization and plasma temperature ($kT_1$) were free in the fit.

The X-ray power-law component (\texttt{pow} model in \texttt{spex}) is used to model the observational data in the hard X-rays,
which is usually thought to be produced by an optically-thin hot corona.
In the fit, its high-energy exponential cut-off ($E_{\rm cut}$) was fixed to 150 keV that is the average value of a FRII-dominated sample \citep{Kang2020}. The low-energy exponential cut-off in the fit was fixed to 13.6 eV. 

The X-ray power-law component is usually accompanied by a neutral X-ray reflection component (\texttt{refl} model in \texttt{spex}) that may reprocess the incident power-law continuum to produce the observed Fe K$\alpha$ line and the Compton hump at higher energies.
In the fit, normalization and photon index of the incident power-law were coupled to those of the \texttt{pow} model, 
and only the reflection scale ($s$) kept free.

\begin{figure}[t!]
\center
\includegraphics[width=\linewidth, clip]{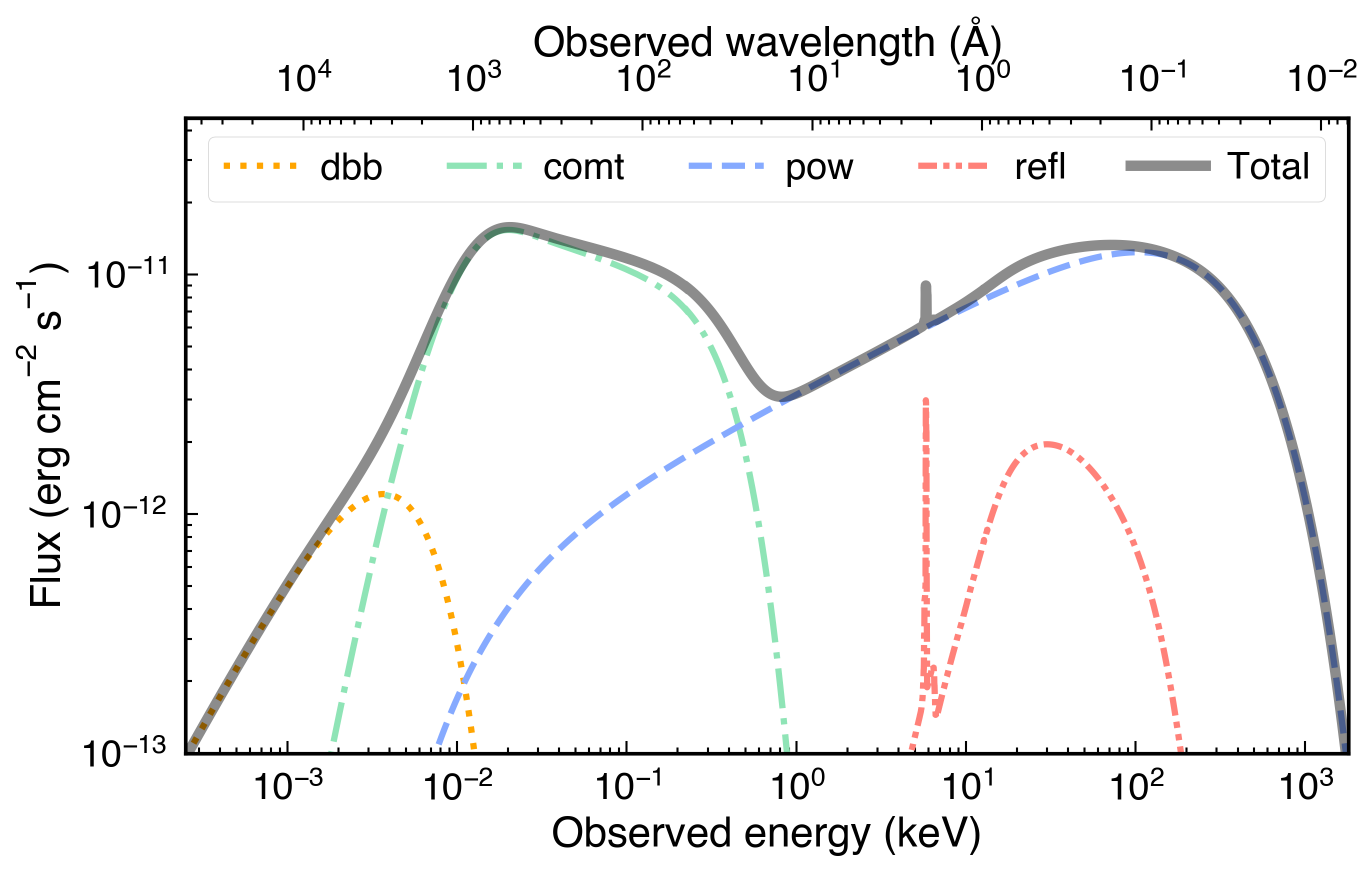}
\caption{The best-fit intrinsic SED models.
The dotted orange line represents an accretion disk blackbody component (\texttt{dbb}).
The dash-dotted green line denotes a warm Comptonization component (\texttt{comt}).
The dashed blue line means an X-ray power-law component (\texttt{pow}).
The dash-dotted red line shows a neutral X-ray reflection component (\texttt{refl}).
The solid black line represents the total best-fit continuum model.
\label{fig:broadbandSED}}
\end{figure}

\subsection{The Galactic neutral gas}
The absorption features caused by the Galactic neutral gas were modeled by the \texttt{hot} model in \texttt{spex}.
Its hydrogen column density ($N_{\rm H}$) was fixed to $6.59\times 10^{20}\ \rm{cm}^{-2}$ \citep[][\citetalias{Mehdipour2019}]{VeronCetty2010}, and its electron temperature was fixed to 0.001 eV to mimic the transmission of a neutral plasma\footnote{See details in the \texttt{spex} manual: \url{https://spex-xray.github.io/spex-help/models/hot.html}}.

\begin{figure*}[t!]
\center
\includegraphics[width=\linewidth, clip]{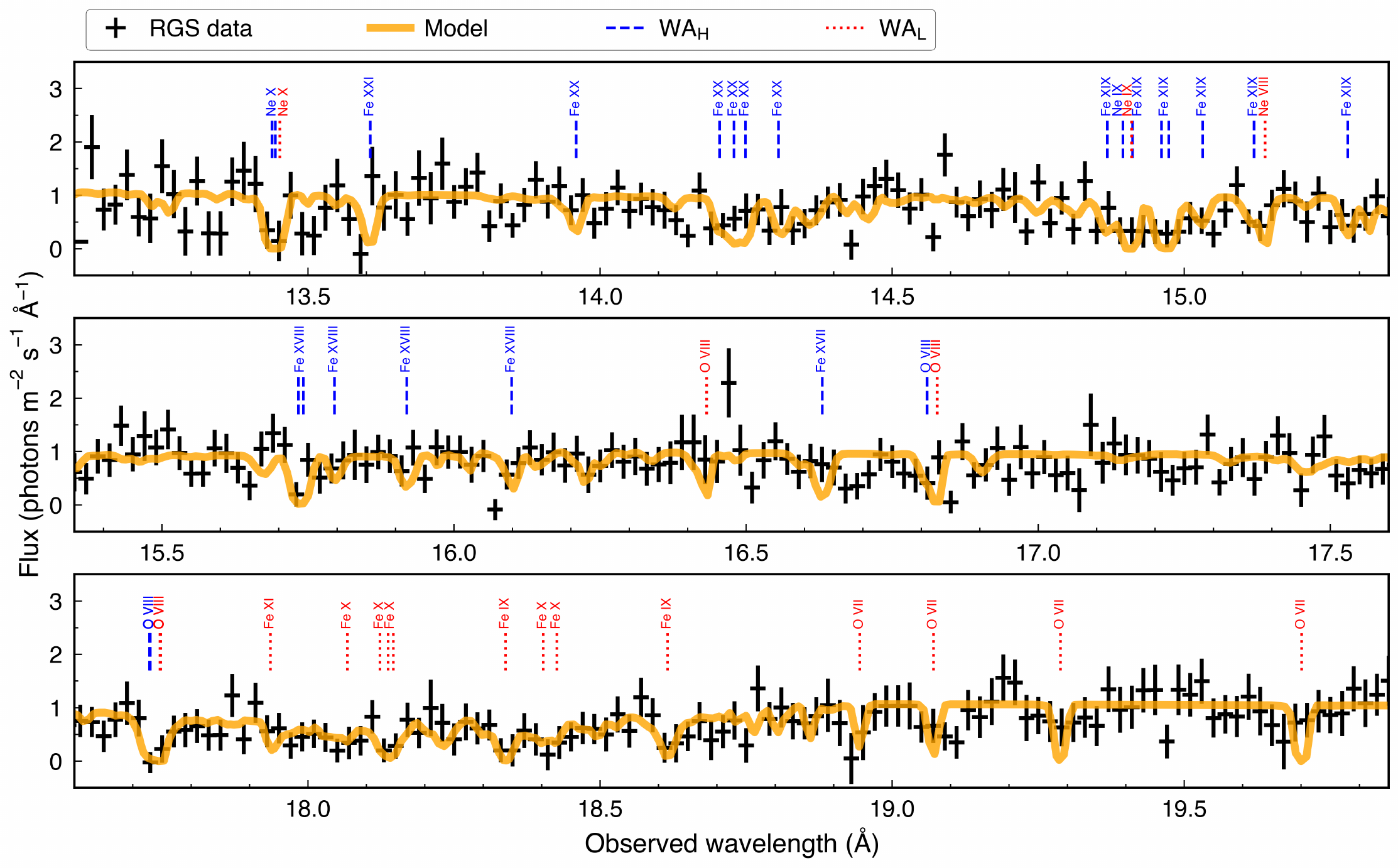}
\caption{The 13--20 {\AA} RGS spectrum with main absorption features produced by the warm absorbers of 3C~59.
The black points represent the RGS spectrum and the solid orange line denotes the best-fit model.
The dashed blue lines and dotted red lines label the main absorption lines produced by the highly-ionized warm absorber (WA$_{\rm H}$)
and the lowly-ionized warm absorber (WA$_{\rm L}$), respectively.
\label{fig:RGSspectradetails}}
\end{figure*}

\subsection{Warm absorbers}
The absorption features produced by the warm absorbers were modeled by the \texttt{pion} model in \texttt{spex}.
The \texttt{pion} model is a robust photoionization code where the photoionization equilibrium can be calculated
self-consistently using the available plasma routines and atomic database of \texttt{spex}.
The hydrogen column density ($N_{\rm H}$), ionization parameter ($\xi$), turbulent velocity ($\sigma_{\rm v}$), and outflowing velocity ($v_{\rm out}$) of this model were free in the fit.
We assume that the X-ray emitting region is fully covered by warm absorbers, 
so the covering factors of warm absorbers are fixed to 1.
The ionization parameter is defined by 
\begin{equation}
\xi = \frac{L_{\rm ion}}{n_{\rm H} r^2},
\label{eq:xi}
\end{equation}
where $L_{\rm ion}$ is the ionizing luminosity over 1--1000 Ryd (13.6 eV to 13.6 keV),
$n_{\rm H}$ is the hydrogen number density of the absorbing clouds,
and $r$ is the radial distance of the absorbing clouds to the central BH \citep{Tarter1969}.

\section{Results and discussions}
\label{sec:resultsdiscussion}

\begin{figure}[h!]
\center
\includegraphics[width=\linewidth, clip]{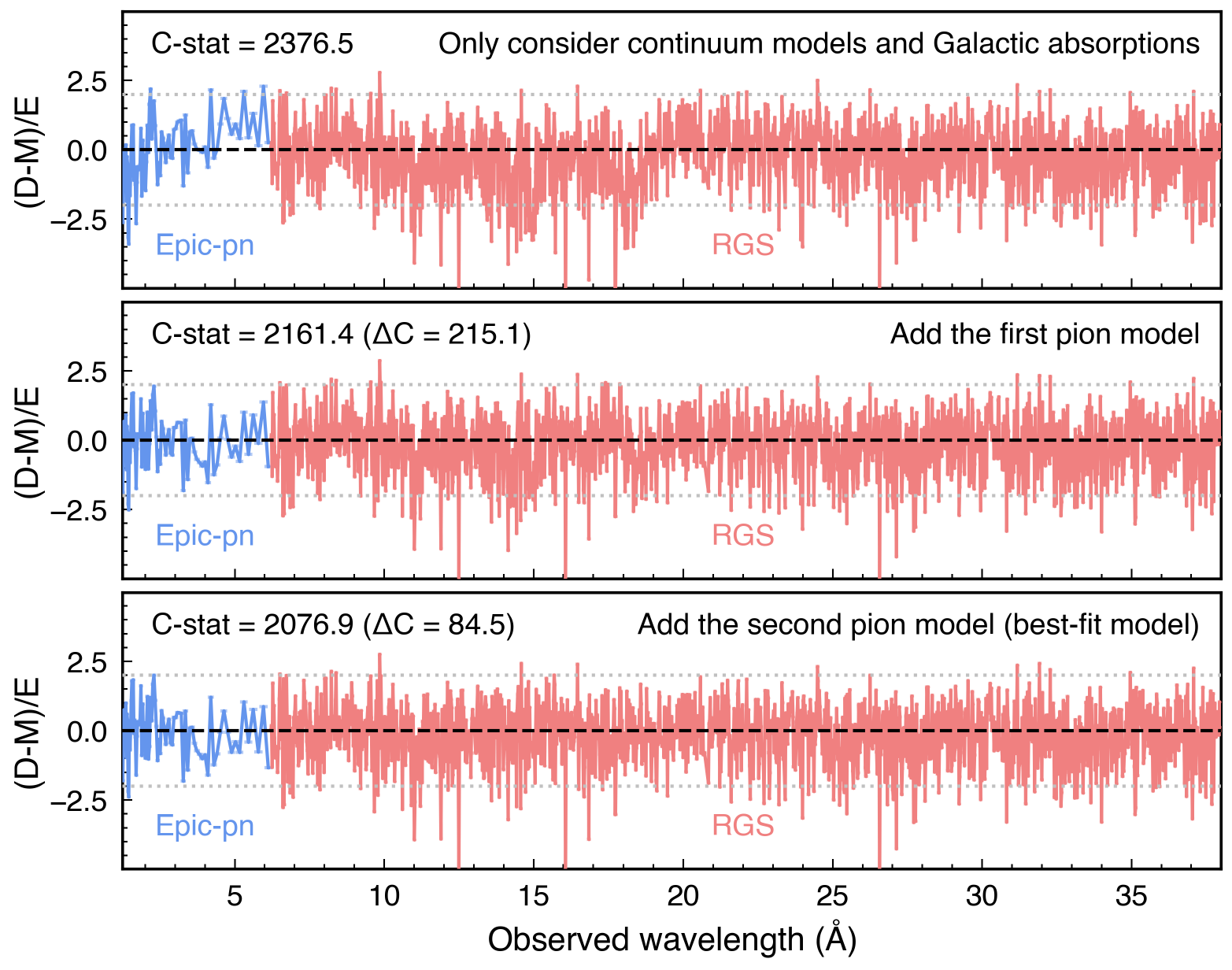}
\caption{Residual of spectral fitting with different models.
The residual ``(D-M)/E'' means (data$-$model)$/$error.
Top panel: The spectral fitting with only continuum models and the Galactic absorptions. 
Middle panel: The spectral fitting after adding the first \texttt{pion} model on the models shown in the top panel.
This \texttt{pion} model is used to model the highly-ionized warm absorber.
The $\Delta C$ here means the C-stat differential between the spectral fitting results in the top and middle panels.
Bottom panel: The spectral fitting after adding the second \texttt{pion} model on the models shown in the middle panel.
This \texttt{pion} model is used to model the lowly-ionized warm absorber.
The $\Delta C$ here means the C-stat differential between the spectral fitting results in the middle and bottom panels.
\label{fig:fittingresidual}}
\end{figure}

\subsection{Intrinsic continuum models of 3C~59}
\label{sec:continuumresults}
The best-fit model can well describe the X-ray spectral data (see Fig. \ref{fig:dataandmodel}).
The unabsorbed intrinsic continuum models are shown in Fig. \ref{fig:broadbandSED} and their parameters are summarized in Table \ref{tab:bestfitparameters}.
The intrinsic luminosities estimated by the total intrinsic continuum model are also summarized in Table \ref{tab:bestfitparameters}.
The intrinsic bolometric luminosity and 0.3--10 keV luminosity of 3C~59 are 
about $L_{\rm bol} = 3.45\times 10^{45}$ erg s$^{-1}$ and $L_{\rm 0.3-10\ keV} = 5.19\times 10^{44}$ erg s$^{-1}$, respectively,
both of which are consistent with those obtained in \citetalias{Mehdipour2019}.
The intrinsic 2--10 keV luminosity is around $L_{\rm 2-10\ keV} = 2.82\times 10^{44}$ erg s$^{-1}$, and
the ionizing luminosity over 1--1000 Ryd is about $L_{\rm ion} = 1.79\times 10^{45}$ erg s$^{-1}$. 
The Eddington luminosity ($L_{\rm Edd}$) of 3C~59 is $1.02 \times 10^{47}$ erg s$^{-1}$, which is estimated by $L_{\rm Edd} = 1.25 \times 10^{38} \times (M_{\rm BH}/M_\odot)$ \citep{Rybicki1979} where $M_{\rm BH}$ is the black hole mass of $10^{8.9}\ M_\odot$ for 3C~59 \citep{Wu2004}.
Thus, according to the definition of Eddington ratio ($\lambda_{\rm Edd} = L_{\rm bol}/L_{\rm Edd}$), 
$\lambda_{\rm Edd}$ of 3C~59 is estimated to be 0.03.
The mass accretion rate $\dot{M}_{\rm acc}$ is 0.62 $M_\odot\ \rm{yr}^{-1}$, which is estimated by $\dot{M}_{\rm acc} \approx L_{\rm bol}/(\eta c^2)$ assuming an accretion efficiency $\eta = 0.1$.
These parameters are summarized in Table \ref{tab:totalproperties}.

\subsection{Properties of warm absorbers in 3C~59}
\label{sec:WAproperties}
The main absorption features produced by warm absorbers of 3C~59 are shown in Fig. \ref{fig:RGSspectradetails}.
Compared to the spectral fitting result with only continuum models and the Galactic absorptions (see the upper panel of Fig. \ref{fig:fittingresidual}), adding the highly-ionized warm absorber (WA$_{\rm H}$) improves the fitting with a C-stat differential of $\Delta C \sim 215.1$ (see the middle panel of Fig. \ref{fig:fittingresidual}).
Then we added a lowly-ionized warm absorber (WA$_{\rm L}$), which improves the fitting with $\Delta C \sim 84.5$ (see the bottom panel of Fig. \ref{fig:fittingresidual}).
Adding the third \texttt{pion} model that is used to model the third warm absorber with lower ionization parameter ($\log [\xi/(\rm{erg}\ \rm{cm}\ \rm{s}^{-1})] < 1$)
only improves the fitting by $\Delta C \sim 10.4$, 
which is insignificant (see the fitting results with three \texttt{pion} models in Section \ref{sec:fit3pion} and Table \ref{tab:threeWAfitpara}) and may require a spectrum with higher S/N to be reliably resolved.
Thus, we only take two warm absorbers (WA$_{\rm H}$ and WA$_{\rm L}$; see Table \ref{tab:bestfitparameters}) into account in this work.
In addition, absorption residual around 9 keV is not significant enough to prove a UFO \citep{Tombesi2014} and thus will not affect the spectroscopy results, such as intrinsic luminosity and ionization balance of warm absorbers.

The best-fit physical parameters for WA$_{\rm H}$ and WA$_{\rm L}$ are summarized in Table \ref{tab:bestfitparameters}.
The WA$_{\rm H}$ has an ionization parameter of $\log [\xi/(\rm{erg}\ \rm{cm}\ \rm{s}^{-1})] = 2.65^{+0.10}_{-0.09}$,
a hydrogen column density of $N_{\rm H} = 0.69^{+0.31}_{-0.17}\times 10^{22}\ \rm{cm}^{-2}$,
and an outflowing velocity of $v_{\rm out} = -528^{+163}_{-222}\ \rm{km}\ \rm{s}^{-1}$.
For WA$_{\rm L}$, these parameters are relatively lower,
which are $\log [\xi/(\rm{erg}\ \rm{cm}\ \rm{s}^{-1})] = 1.65\pm 0.11$,
$N_{\rm H} = 0.31^{+0.04}_{-0.03}\times 10^{22}\ \rm{cm}^{-2}$,
and $v_{\rm out} = -228^{+121}_{-122}\ \rm{km}\ \rm{s}^{-1}$.
Parameter contour maps of WA$_{\rm H}$ and WA$_{\rm L}$ are shown in Fig. \ref{fig:contourmaps}.
The WA$_{\rm H}$ mainly produces absorption features of Fe XVIII--XXI lines over 13--17 {\AA} (observer frame),
and it also produces the absorption lines of O VIII and Ne IX--X
(see Fig. \ref{fig:RGSspectradetails}).
The main absorption features produced by WA$_{\rm L}$ include Fe IX--XI and O VII--VIII over 17--20 {\AA} (observer frame),
and other highly-ionized lines like Ne VIII--X (see Fig. \ref{fig:RGSspectradetails}).
The velocity profiles for six of the above-mentioned absorption lines are shown in Fig. \ref{fig:velocityprofile}.

\subsection{Radial location of warm absorbers in 3C~59}
\label{sec:distanceWA}

\begin{figure}[t!]
\center
\includegraphics[width=\linewidth, clip]{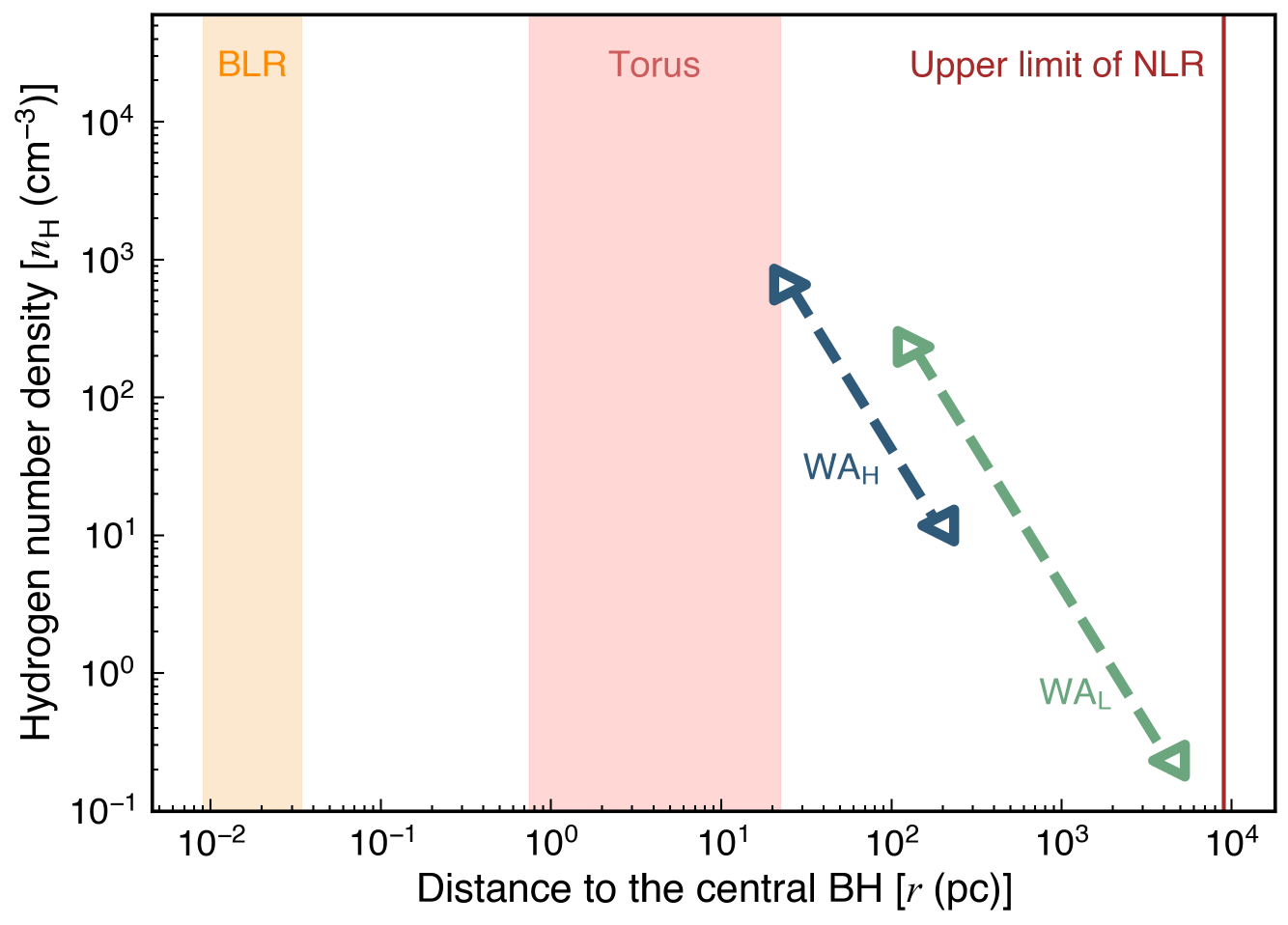}
\caption{Distance of warm absorbers to the central BH in 3C~59.
The solid blue and green lines represent the $n_{\rm H}$--$r$ distributions for WA$_{\rm H}$ and WA$_{\rm L}$, respectively.
The markers ``$\rhd$'' and ``$\lhd$'' denote the lower and upper limits of $r$, respectively. 
The orange region means the size of broad (emission-)line region in 3C~59.
The red region shows the size of torus in 3C~59.
The dotted brown line represents the maximum distance of narrow (emission-)line region to the BH in 3C~59.
We refer readers to Section \ref{sec:distanceWA} for detailed calculation of these sizes.
\label{fig:distance}}
\end{figure}

We separately estimated the upper and lower limits for the distance ($r$) of warm absorbers to the central BH.
The upper limit is estimated based on the assumption that 
the thickness of the absorber cloud ($\Delta r$) does not exceed its distance to the BH \citep{Krolik2001,Blustin2005}.
Considering that $N_{\rm H} \approx n_{\rm H} \Delta r$, 
the upper limit of the distance $r_{\rm max} \approx \Delta r \approx N_{\rm H}/n_{\rm H}$.
Combining with Eq. \ref{eq:xi}, $r_{\rm max}$ can be expressed as $r_{\rm max} \approx L_{\rm ion}/(\xi N_{\rm H})$.
The lower limit of the distance is estimated based on the assumption that
the outflowing velocities ($v_{\rm out}$) of warm absorbers should be higher than the 
escape velocities $v_{\rm esc} = \sqrt{2 G M_{\rm BH} / r}$ where $G$ is the gravitational constant \citep{Blustin2005}.
Thus, the lower limit of the distance is estimated by $r_{\rm min} = 2 G M_{\rm BH}/v_{\rm out}^2$.
According to these definitions, the distance is 25 pc $\lesssim r \lesssim$ 188 pc for $\rm{WA}_{\rm H}$
and 135 pc $\lesssim r \lesssim$ 4 kpc for $\rm{WA}_{\rm L}$ (see Table \ref{tab:totalproperties} and Fig. \ref{fig:distance}).
Here we want to stress that $r_{\rm min}$ and $r_{\rm max}$ indicate a rough distance range 
rather than a true physical scale ($\Delta r$) of warm absorbers.
True $\Delta r$ may be much smaller than its distance to the BH \citep{Sadaula2023},
which is difficult to estimate.

In Fig. \ref{fig:distance}, we also show the positions of two warm absorbers relative to the broad (emission-)line region (BLR),
dust torus, and NLR.
The distance of the BLR to the central BH is estimated by its correlation with AGN continuum luminosity at rest-frame 5100\ \AA\ ($\lambda L_{\lambda} (\rm 5100\ \AA)$),
which is $\log (R_{\rm BLR}/{\rm ld}) = (1.69 \pm 0.23) + (0.37^{+0.18}_{-0.17})\times \log [\lambda L_{\lambda} (\rm 5100\ \AA) /(10^{44}\ \rm{erg}\ \rm{s}^{-1})]$
\citep{GRAVITY2024}.
The $\lambda L_{\lambda} (\rm 5100\ \AA)$ of 3C~59 is given by the best-fit total intrinsic continuum model, which is about $3.51\times 10^{43}\ \rm{erg}\ \rm{s}^{-1}$.
Here we do not use the LAMOST spectrum to estimate $\lambda L_{\lambda} (\rm 5100\ \AA)$, because the host galaxy of 3C~59 has a non-negligible contribution at rest-frame 5100 {\AA}, which is difficult to reliably deduct (see details in Section \ref{sec:continuummodel}).
The $R_{\rm BLR}$ of 3C~59 is around 0.009--0.03 pc.
The inner radius of dust torus is estimated by $R_{\rm torus, in} = 0.4\times (L_{\rm bol}/10^{45}\ \rm{erg}\ \rm{s}^{-1})^{0.5}\times (1500\ K/T_{\rm d})^{2.6}$ pc with a dust temperature $T_{\rm d} = 1500$ K \citep{Nenkova2008}, and the outer radius of dust torus is assumed to be $R_{\rm torus, out} \approx 30R_{\rm torus, in}$ \citep[e.g.,][]{Nenkova2008,Zhou2024}.
For 3C~59, $R_{\rm torus, in}$ and $R_{\rm torus, out}$ is about 0.7 pc and 22 pc, respectively.
The distance of the NLR is estimated by its correlation with [OIII] $\lambda$5007 \AA\ line luminosity ($L_{[\rm OIII]}$),
which is $\log (R_{\rm NLR}/{\rm pc}) =  (0.250 \pm 0.018)\times \log [L_{[\rm OIII]} /(10^{42}\ \rm{erg}\ \rm{s}^{-1})] + (3.746 \pm 0.028)$ \citep{Liu2013}.
The $L_{[\rm OIII]}$ is estimated through the spectral fitting with the LAMOST spectrum,
which is $6.86\times 10^{42}\ \rm{erg}\ \rm{s}^{-1}$ (see details in Appendix \ref{sec:LAMOSTspectralfitting}).
We want to stress that this $L_{[\rm OIII]}$ also includes star-formation-induced contribution that is difficult to be separated from the AGN contributions, so here we only used this $L_{[\rm OIII]}$ as an upper limit to estimate the maximum distance of NLR, which is $R_{\rm NLR,max} \sim 9$ kpc.
As Fig. \ref{fig:distance} shows, warm absorbers of 3C~59 are located between the outer torus and NLR.
The distances of BLR, dust torus, and NLR are summarized in Table \ref{tab:totalproperties}.

\subsection{Comparison between different spectral fitting strategies}
\label{sec:fittingstrategy}
\citetalias{Mehdipour2019} also found two warm absorbers for 3C~59
(WA$_{\rm H}$: $N_{\rm H} = (0.81\pm 0.08) \times 10^{22}\ \rm{cm}^{-2}$, 
$\log[\xi/(\rm{erg}\ \rm{cm}\ \rm{s}^{-1})] = 2.42 \pm 0.03$, 
and $v_{\rm out} = -1000\pm 120\ \rm{km}\ \rm{s}^{-1}$;
WA$_{\rm L}$: $N_{\rm H} = (0.40\pm 0.02) \times 10^{22}\ \rm{cm}^{-2}$, 
$\log[\xi/(\rm{erg}\ \rm{cm}\ \rm{s}^{-1})] = 1.20 \pm 0.05$, 
and $v_{\rm out} = -3530\pm 130\ \rm{km}\ \rm{s}^{-1}$).
We got significantly different results comparing to \citetalias{Mehdipour2019}.
There are several main differences between our analysis and \citetalias{Mehdipour2019}:
1) we additionally used NIR-opt-UV data to constrain the continuum model comparing to \citetalias{Mehdipour2019};
2) \citetalias{Mehdipour2019} included EPIC-pn data below 2 keV, while we excluded them;
3) \citetalias{Mehdipour2019} used an accretion disk model (\texttt{mbb}) to describe the emission from optical to soft X-ray bands, 
while we used both a warm Comptonization component (\texttt{comt}) and a disk model (\texttt{dbb}) to collectively describe these emissions.

The comparison between our results and \citetalias{Mehdipour2019} also reminds us that 
different spectral fitting strategies may also have a potential impact on the best-fit result,
such as whether to incorporate NIR-optical-UV data to constrain the continuum models, the choice of spectral coverage of each instrument, and the composition of SED models.
The existence of this issue calls for rigorous caution when comparing or combining results from different works.
However, until now, this issue is still lack of adequate studies.
To check the potential impact of different spectral fitting strategies on the best-fit results,
we perform the following two spectral fitting tests: 
(1) spectral fitting without NIR-optical-UV data and adopting different SED compositions (Test 1);
(2) spectral fitting with inclusion of the EPIC-pn spectrum in the energy range of 0.3--2 keV (Test 2).

\begin{figure}[t!]
\center
\includegraphics[width=\linewidth, clip]{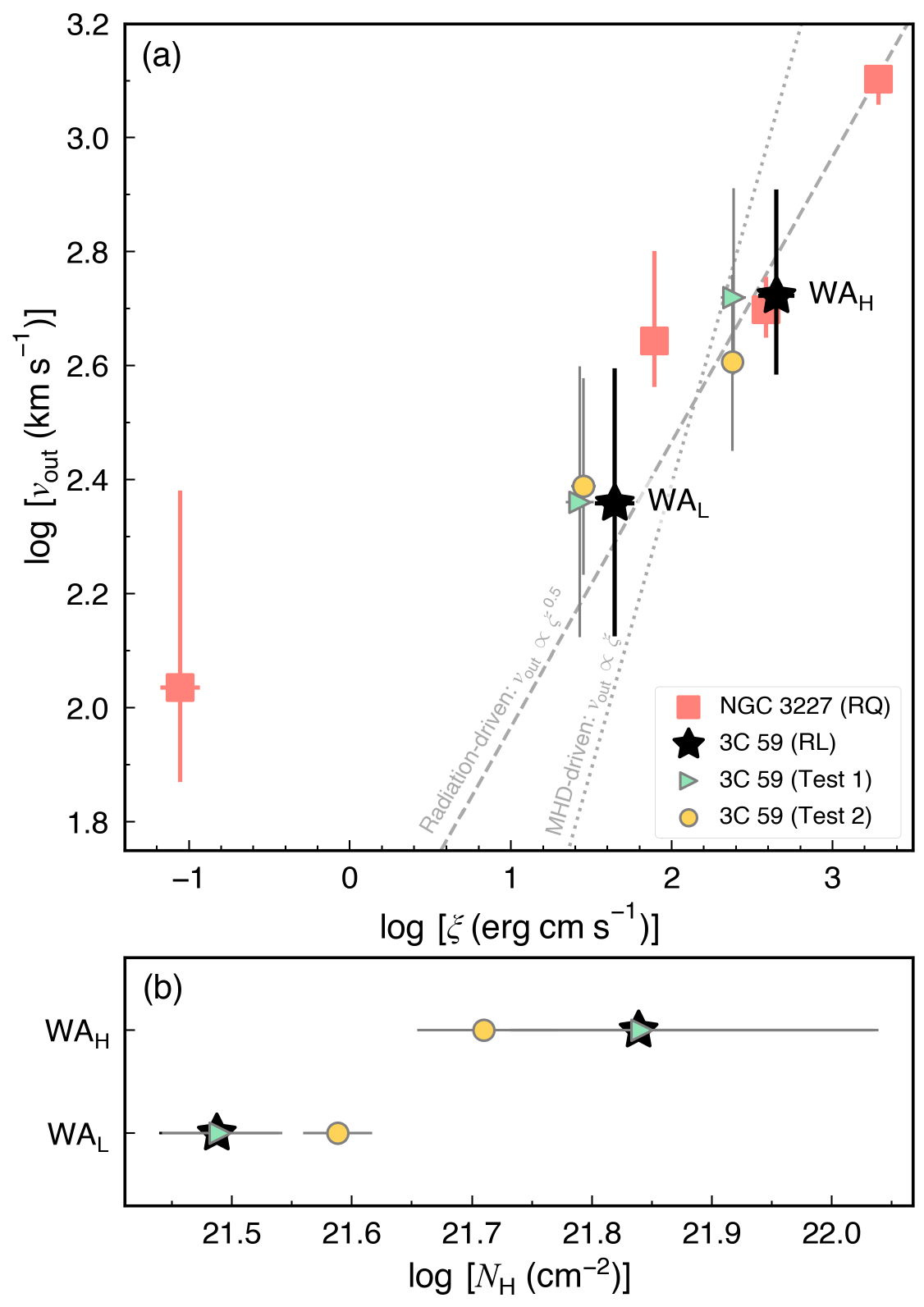}
\caption{Outflowing velocity ($v_{\rm out}$) plotted against ionization parameter ($\xi$) (panel (a))
and distribution of hydrogen column density (panel (b)) of a radio-loud (RL) AGN 3C 59.
The black stars are the best-fit results obtained by the spectral fitting strategy that we adopted in this work.
The symbols with gray edges represent the best-fit results based on different spectral fitting strategies: spectral fitting without NIR-optical-UV data and adopting different SED compositions (Test 1; green triangles), and spectral fitting with inclusion of the EPIC-pn spectrum in the energy range of 0.3--2 keV (Test 2; yellow circles) (see details in Section \ref{sec:fittingstrategy}).
The red squares are the best-fit results for a radio-quiet (RQ) AGN NGC~3227, which are derived from \cite{Wang2022a}.
The $v_{\rm out}$-$\xi$ relations of the radiation-pressure-driven (dashed gray line) and 
MHD-driven (dotted gray line) mechanisms are scaled to the average values of WA$_{\rm H}$ and WA$_{\rm L}$ of 3C~59 (two black stars).
\label{fig:scalingrelation}}
\end{figure}

The spectral fitting strategy of Test 1 differs from that adopted in this work (see details in Section \ref{sec:spectralfitSED})
primarily in the exclusion/inclusion of NIR-optical-UV data and the composition of broadband SED.
In this test, we did not utilize NIR-optical-UV data and adopted the same SED composition (\texttt{mbb}, \texttt{pow}, and \texttt{refl}) as \citetalias{Mehdipour2019}.
The \texttt{mbb} model is used to describe the emission from optical to soft X-rays bands.
The best-fit parameters of warm absorbers of Test 1 are shown as the green triangles in Fig. \ref{fig:scalingrelation}.
Comparing Test 1 with the spectral fitting strategy adopted in this work (black stars in Fig. \ref{fig:scalingrelation}),
ionization parameters of warm absorbers show a difference.
At the 95\% confidence level, this difference is statistically significant for WA$_{\rm H}$ and is not statistically significant for WA$_{\rm L}$. This difference may be mainly due to the different ionizing luminosity caused by the different SED composition/shape (the ionizing luminosity given by Test 1 is about 50\% of the value shown in Table \ref{tab:bestfitparameters}).
Further, we found that the AGN fluxes at $g$, $r$, and $z$ filters of DESI
are about 50 times higher than the values predicted by the disk model in Test 1,
which indicates that two SED components are essential to fully describe the emission from optical to soft X-ray bands.
In conclusion, NIR-optical-UV data are very essential for the construction of broadband SED,
which may further affect the estimation of physical properties of warm absorbers.

The spectral fitting strategy of Test 2 differs from that adopted in this work
primarily in the inclusion/exclusion of EPIC-pn spectrum in the energy range of 0.3--2 keV during spectral fitting.
In Test 2, we adopted the same broadband SED composition as that in Section \ref{sec:spectralfitSED}.
The best-fit parameters of warm absorbers of Test 2 are shown as the yellow circles in Fig. \ref{fig:scalingrelation}.
Comparing Test 2 with the spectral fitting strategy adopted in this work (black stars in Fig. \ref{fig:scalingrelation}),
ionization parameters and hydrogen column densities of warm absorbers show differences.
At the 95\% confidence level, the difference of ionization parameter is statistically significant for WA$_{\rm H}$ and is not statistically significant for WA$_{\rm L}$,
while the differences of hydrogen column density are not statistically significant for both WA$_{\rm H}$ and WA$_{\rm L}$.
These differences are mainly due to the different SED shape.

\subsection{Driving mechanisms of warm absorbers in 3C~59 and the comparison with radio-quiet AGN NGC 3227}
The correlation between outflowing velocity and ionization parameter
can trace the driving mechanisms of outflows \citep{Tombesi2013}.
Based on the assumption that outflow momentum rate is comparable to the momentum rate of the radiation field \citep[e.g.,][]{Gofford2015},
radiation-pressure-driven mechanism predicts that $v_{\rm out} \varpropto \xi^{0.5}$ \citep{Tombesi2013}.
According to one of the most profound scaling relations for magnetohydrodynamic (MHD) winds \citep[e.g.,][]{KoniglA2000,PelletierG1992}, which shows that outflow rate of MHD wind is proportional to the accretion rate,
MHD-driven mechanism predicts that $v_{\rm out} \varpropto \xi$ \citep{Tombesi2013}.
As Fig. \ref{fig:scalingrelation} shows, 
warm absorbers of 3C~59 appear to be more consistent with the radiation-pressure-driven mechanism than MHD-driven mechanism.
Warm absorbers may be mainly dusty winds, for which the radiation-pressure-driven mechanism usually plays a crucial role \citep{Yamada2024}.
However, \cite{Wang2022} found that neither radiation-pressure-driven nor MHD-driven mechanisms 
can fully explain the $v_{\rm out}$-$\xi$ relation of warm absorbers in the radio-quiet AGNs, 
especially for the warm absorbers with $\log [\xi/(\rm{erg}\ \rm{cm}\ \rm{s}^{-1})] < 1$.  
They argued that the pressure equilibrium between radiation pressure on the warm absorbers and drag pressure from the ACM can well explain the $v_{\rm out}$-$\xi$ relation,
and they further used the index of $v_{\rm out}$-$\xi$ relation to deduce the density profile of ACM,
which implies that the accretion physics in these radio-quiet AGNs follow the standard thin disk model.
However, we do not find any reliable warm absorbers with $\log [\xi/(\rm{erg}\ \rm{cm}\ \rm{s}^{-1})] < 1$ for 3C~59,
so we cannot make detailed statistical analysis for ACM density profile in 3C~59.
In the future, we will require more high-resolution X-ray spectra with longer exposure to make further analysis.

To compare the driven mechanisms of warm absorber outflows between radio-quiet and radio-loud AGNs,
we plot the results of a radio-quiet AGN NGC 3227 from \cite{Wang2022a} in Fig. \ref{fig:scalingrelation}.
\cite{Wang2022a} adopted the same spectral fitting strategy for NGC 3227 as that for 3C 59 in this work.
For warm absorbers with $\log[\xi/(\rm{erg}\ \rm{cm}\ \rm{s}^{-1})] > 1$, 
3C~59 has a generally similar $v_{\rm out}$-$\xi$ relation to NGC~3227 (see Fig. \ref{fig:scalingrelation}). 
It demonstrates that at least for warm absorbers with $\log[\xi/(\rm{erg}\ \rm{cm}\ \rm{s}^{-1})] > 1$,
radio-quiet AGN NGC~3227 and radio-loud AGN 3C~59 appear to have similar driving mechanisms,
and jets may play a negligible role in these processes.
However, due to limited sample size, we cannot extend this conclusion to all the radio-quiet and radio-loud AGNs.
Even so, this work present a comparative analysis of outflow driven mechanisms between radio-loud and radio-quiet AGNs.
Serving as a pilot study, this work also validates our spectral fitting strategy that will be applied in more radio-quiet and radio-loud AGNs to draw a more statistically robust conclusion.

   \begin{table}[t!]
   \caption{Properties of 3C~59 \label{tab:totalproperties}}
   \centering
   \begin{tabular}{rlccccccccccccc}
   \hline\hline\xrowht[()]{6pt}
    & Source properties & Ref. \\ 
   \hline
   (1) & $M_{\rm BH} = 10^{8.9}\ M_\odot$ & \cite{Wu2004} \\
   (2) & $L_{\rm bol} = 3.45\times 10^{45}$ erg s$^{-1}$ & Section \ref{sec:continuumresults} \\
   (3) & $L_{\rm Edd} = 1.02\times 10^{47}$ erg s$^{-1}$ & Section \ref{sec:continuumresults} \\
   (4) & $\lambda_{\rm Edd} = 0.03$ & Section \ref{sec:continuumresults} \\
   (5) & $L_{\rm ion} = 1.79\times 10^{45}$ erg s$^{-1}$ & Section \ref{sec:continuumresults} \\
   (6) & $\dot{M}_{\rm acc} = 0.62\ M_\odot\ \rm{yr}^{-1}$ & Section \ref{sec:continuumresults} \\
   (7) & 0.009 pc $\lesssim R_{\rm BLR} \lesssim$ 0.03 pc & Section \ref{sec:distanceWA} \\
   (8) & 0.7 pc $\lesssim R_{\rm torus} \lesssim$ 22 pc & Section \ref{sec:distanceWA} \\
   (9) & $R_{\rm NLR} \lesssim$ 9 kpc & Section \ref{sec:distanceWA} \\
   \hline\xrowht[()]{5pt}
   & $\rm{WA}_{\rm H}$ properties & Ref. \\ 
   \hline
   (10) & 25 pc $\lesssim r \lesssim$ 188 pc & Section \ref{sec:distanceWA} \\
   (11) & $\dot{P}_{\rm abs} = 1.28\times 10^{35}\ \rm{erg}\ \rm{m}^{-1}$ & Section \ref{sec:massoutflowrate} \\ 
   (12) & $\dot{P}_{\rm scatt} = 2.73\times 10^{34}\ \rm{erg}\ \rm{m}^{-1}$ & Section \ref{sec:massoutflowrate} \\
   (13) & $\dot{M}_{\rm out} = 0.47\ M_\odot\ \rm{yr}^{-1}$ & Section \ref{sec:massoutflowrate} \\
   (14) & $\dot{E}_{\rm K} = 4.11 \times 10^{40}\ \rm{erg}\ \rm{s}^{-1}$ & Section \ref{sec:massoutflowrate} \\
   \hline\xrowht[()]{5pt}
   & $\rm{WA}_{\rm L}$ properties & Ref. \\ 
   \hline
   (15) & 135 pc $\lesssim r \lesssim$ 4 kpc & Section \ref{sec:distanceWA} \\
   (16) & $\dot{P}_{\rm abs} = 4.43\times 10^{35}\ \rm{erg}\ \rm{m}^{-1}$ & Section \ref{sec:massoutflowrate} \\ 
   (17) & $\dot{P}_{\rm scatt} = 1.22\times 10^{34}\ \rm{erg}\ \rm{m}^{-1}$ & Section \ref{sec:massoutflowrate} \\
   (18) & $\dot{M}_{\rm out} = 3.15\ M_\odot\ \rm{yr}^{-1}$ & Section \ref{sec:massoutflowrate} \\
   (19) & $\dot{E}_{\rm K} = 5.20 \times 10^{40}\ \rm{erg}\ \rm{s}^{-1}$ & Section \ref{sec:massoutflowrate} \\
   \hline
   \end{tabular}
   \tablefoot{(1): black hole mass. (2): bolometric luminosity. (3): Eddington luminosity. 
   (4): Eddington ratio. (5): 1--1000 Ryd ionizing luminosity.
   (6): mass accretion rate. (7): distance of BLR to BH. (8): distance of dust torus to BH. (9): distance of NLR to BH.
   (10) and (15): distance of warm absorbers to BH. 
   (11) and (16): momentum of radiation being absorbed by warm absorbers. 
   (12) and (17): momentum of radiation being scattered by warm absorbers.
   (13) and (18): mass outflow rate of warm absorbers. 
   (14) and (19): kinetic energy of warm absorbers.
   }
   \end{table}

\subsection{Mass outflow rate of warm absorbers in 3C~59}
\label{sec:massoutflowrate}
The mass outflow rate of warm absorbers can be estimated by $\dot{M}_{\rm out}\sim (\dot{P}_{\rm abs} + \dot{P}_{\rm scatt}) / v_{\rm out}$,
where $\dot{P}_{\rm abs}$ is the momentum of radiation being absorbed by warm absorbers 
and $\dot{P}_{\rm scatt}$ is the momentum of radiation being scattered by warm absorbers \citep{Blustin2005}.
Here, $\dot{P}_{\rm abs}$ is calculated by $\dot{P}_{\rm abs}=L_{\rm abs}/c$, where $L_{\rm abs}$ is the luminosity absorbed by the warm absorber over 1--1000 Ryd, given by the spectral fitting.
The $\dot{P}_{\rm scatt}$ is estimated by $\dot{P}_{\rm scatt}=L_{\rm ion} (1 - e^{-\tau_{\rm T}}) / c$, where $\tau_{\rm T}$ is the optical depth for Thomson scattering, given by $\tau_{\rm T}=\sigma_{\rm T} N_{\rm H}$ ($\sigma_{\rm T}$ is the Thomson cross-section).
The detailed values of $\dot{P}_{\rm abs}$ and $\dot{P}_{\rm scatt}$ are summarized in Table \ref{tab:totalproperties}.
The mass outflow rate is around 0.47 $M_\odot\ \rm{yr}^{-1}$ and 3.15 $M_\odot\ \rm{yr}^{-1}$ for $\rm{WA}_{\rm H}$ and $\rm{WA}_{\rm L}$, respectively.
The kinetic energy of warm absorbers can be estimated by $\dot{E}_{\rm K} = 1/2 \dot{M}_{\rm out} v_{\rm out}^2$, 
which is $4.11 \times 10^{40}\ \rm{erg}\ \rm{s}^{-1}$ for $\rm{WA}_{\rm H}$ and $5.20 \times 10^{40}\ \rm{erg}\ \rm{s}^{-1}$ for $\rm{WA}_{\rm L}$.
\cite{Hopkins2010} found that if $\dot{E}_{\rm K}/L_{\rm bol}$ of a wind is higher than 0.5\%, this wind will have an efficient feedback to the host galaxy.
However, for both $\rm{WA}_{\rm H}$ and $\rm{WA}_{\rm L}$, $\dot{E}_{\rm K}/L_{\rm bol}$ is $\sim 0.002\%$, 
which implies that warm absorbers in 3C~59 do not have a significant feedback to the host galaxy.

\section{Summary and conclusion}
\label{sec:summary}

Based on the multi-wavelength data from NIR to hard X-ray bands detected by DESI, GALEX, and XMM-Newton,
we used \texttt{spex} code to construct intrinsic broadband SED and make X-ray spectral analysis for a radio-loud AGN 3C~59 (Fig. \ref{fig:dataandmodel} and Fig. \ref{fig:broadbandSED}).

According to the best-fit results, we found two warm absorbers for 3C~59,
which present significantly different physical properties 
(Table \ref{tab:bestfitparameters}).
The highly-ionized warm absorber with an ionization parameter of $\log[\xi/(\rm{erg}\ \rm{cm}\ \rm{s}^{-1})] = 2.65^{+0.10}_{-0.09}$ has a higher column density ($N_{\rm H} = 0.69^{+0.31}_{-0.17}\times 10^{22}\ \rm{cm}^{-2}$) and a higher outflowing velocity ($v_{\rm out} = -528^{+163}_{-222}\ \rm{km}\ \rm{s}^{-1}$) than the lowly-ionized warm absorber ($\log[\xi/(\rm{erg}\ \rm{cm}\ \rm{s}^{-1})] = 1.65\pm 0.11$, $N_{\rm H} = 0.31^{+0.04}_{-0.03}\times 10^{22}\ \rm{cm}^{-2}$, and $v_{\rm out} = -228^{+121}_{-122}\ \rm{km}\ \rm{s}^{-1}$). These warm absorbers are located between outer torus and narrow (emission-)line region (Table \ref{tab:totalproperties} and Fig. \ref{fig:distance}), while they do not show a significant feedback to the host galaxy.

We found that different spectral fitting strategies, such as the inclusion of NIR to UV data, the choice of energy range of spectrum, or the composition of broadband SED, have an impact on the estimations of physical parameters of warm absorbers (Fig. \ref{fig:scalingrelation}). The existence of this issue also calls for rigorous caution when comparing or combining results from different works.

  The positive correlation between $v_{\rm out}$ and $\xi$ of these two warm absorbers
  can be explained by the radiation-pressure-driven mechanism (Fig. \ref{fig:scalingrelation}).
  In addition, this correlation is almost consistent with that in a radio-quiet AGN NGC~3227 
  based on the same spectral fitting strategy,
  which implies that the warm absorber outflows of 3C 59 and NGC 3227 have the same driven mechanisms and jets may play a negligible role in these processes (Fig. \ref{fig:scalingrelation}).
  Following this pilot study, we will apply the same spectral fitting strategy in more radio-loud and radio-quiet AGNs to draw a more statistically robust conclusion.

\begin{acknowledgements}

We thank the anonymous referee for the constructive comments that greatly improved this paper.
The data and scripts used in this work are available at 
Zenodo, DOI: \href{https://doi.org/10.5281/zenodo.17292242}{10.5281/zenodo.17292242}.
This work was supported by National Natural Science Foundation of China 
(Project No.12173017 and Key Project No.12141301), 
National Key R\&D Program of China (grant no. 2023YFA1605600), 
Scientific Research Innovation Capability Support Project for Young Faculty 
(Project No. ZYGXQNJSKYCXNLZCXM-P3), 
and the China Manned Space Project.
Y.J.W. acknowledges support by National Natural Science Foundation of China (Project
No. 12403019) and Jiangsu Natural Science Foundation (Project No. BK20241188).
Z. C. He acknowledges the support of the National Natural Science
Foundation of China (Grant Nos. 12222304, 12192220,
and 12192221).
Y.Q.X. acknowledges support from National Natural Science Foundation
of China (Project No. 12025303 and Project No. 12393814).
Guoshoujing Telescope (the Large Sky Area Multi-Object Fiber Spectroscopic Telescope LAMOST) is a National Major Scientific Project built by the Chinese Academy of Sciences. Funding for the project has been provided by the National Development and Reform Commission. LAMOST is operated and managed by the National Astronomical Observatories, Chinese Academy of Sciences.

\end{acknowledgements}

\bibliographystyle{aa}
\bibliography{ms.bib}

\begin{appendix}

\setcounter{table}{0}
\renewcommand{\thetable}{A\arabic{table}}

\setcounter{figure}{0}
\renewcommand{\thefigure}{A\arabic{figure}}

\section{X-ray spectral analysis with three \texttt{pion} models}
\label{sec:fit3pion}
In Table \ref{tab:threeWAfitpara}, we show the spectral fitting with three \texttt{pion} models.
The spectral fitting improves with $\Delta C \sim 10.4$ 
comparing to that with two \texttt{pion} models (see Table \ref{tab:bestfitparameters}).
The WA$_{\rm H}$ and WA$_{\rm L}$ have similar physical parameters to those in the spectral fitting with two \texttt{pion} models,
and the third \texttt{pion} model may describe another absorber component.
This component shows a much lower column density of $N_{\rm H} \sim 10^{20.2}\ \rm{cm}^{-1}$ 
and a much lower ionization parameter of $\log[\xi/(\rm{erg}\ \rm{cm}\ \rm{s}^{-1})] \sim -0.88$ than WA$_{\rm H}$ and WA$_{\rm L}$. 
Due to the large uncertainties for the outflowing velocity of this component,
we cannot draw a solid conclusion about whether it is another warm absorber or not.
In addition, the spectral fitting improves slightly with the addition of the third \texttt{pion} model,
so we do not take it into account in this work (see Section \ref{sec:WAproperties}).

   \begin{table*}[h!]
   \caption{Best-fit results with three \texttt{pion} models for 3C~59 \label{tab:threeWAfitpara}}
   \centering
   \renewcommand{\arraystretch}{1.2}
   \begin{tabular}{llllccccccccccc}
   \hline\hline\xrowht[()]{6pt}
   Component & Model & Parameter & Symbol & Value \\
   \hline\xrowht[()]{5pt}
   Disk blackbody component & \texttt{dbb} & Normalization &  $A$ ($10^{25}\ \rm{m}^2$) & 1.34 (scaled) \\
    (NIR-optical-UV)  & & Temperature & $kT_{\rm BB}$ (eV) & 3.4 (fixed) \\
   \hline\xrowht[()]{5pt}
   Warm Comptonization component  & \texttt{comt} & Normalization & $A$ ($10^{56}\ \rm{ph} \rm{s}^{-1} \rm{keV}^{-1}$) & $3.24^{+1.08}_{-0.74}$ \\
   (soft X-ray excess)  & & Seed photons temperature & $kT_0$ (eV) & $3.4$ (fixed) \\
    & & Plasma temperature & $kT_1$ (eV) & $93.94^{+4.59}_{-4.42}$ \\
    & & Optical depth & $\tau$ & 30 (fixed) \\
   \hline\xrowht[()]{5pt}
   X-ray power-law component & \texttt{pow} & Normalization & $A$ ($10^{52}\ \rm{ph} \rm{s}^{-1} \rm{keV}^{-1}$) & $5.94^{+0.19}_{-0.15}$\\
           & & Photon index & $\Gamma$ & $1.63 \pm 0.02$ \\
   \hline\xrowht[()]{5pt}
   Neutral reflection component  & \texttt{refl} & Incident power-law normalization & $A$ ($10^{52}\ \rm{ph} \rm{s}^{-1} \rm{keV}^{-1}$) & $5.97$ (coupled) \\
     & & Incident power-law photon index & $\Gamma$ & 1.63 (coupled) \\
    & & Reflection scale & $s$ & $0.21^{+0.06}_{-0.05}$ \\
   \hline\xrowht[()]{5pt}
   Galactic neutral gas & \texttt{hot} & Hydrogen column density & $N_{\rm H}$ ($10^{20}$ cm$^{-2}$) & $6.59$ (fixed) \\
           & & Electron temperature & $kT_{\rm e}$ (eV) & $0.001$ (fixed) \\
   \hline\xrowht[()]{5pt}
   Warm absorber (WA$_{\rm H}$) & \texttt{pion} & Hydrogen column density & $N_{\rm H}$ ($10^{22}$ cm$^{-2}$) & $0.64^{+0.28}_{-0.16}$ \\
           & & $^{10}\log$ of ionization parameter & $\log\ [\xi\ (\rm{erg}\ \rm{cm}\ \rm{s}^{-1})]$ & $2.64\pm 0.09$ \\
           & & Turbulent velocity & $\sigma_v$ & $189^{+99}_{-94}$ \\
           & & Outflowing velocity & $v_{\rm out}\ (\rm{km}\ \rm{s}^{-1})$ & $-553^{+166}_{-204}$ \\
   \hline\xrowht[()]{5pt} 
   Warm absorber (WA$_{\rm L}$) & \texttt{pion} & Hydrogen column density & $N_{\rm H}$ ($10^{22}$ cm$^{-2}$) & $0.31\pm 0.03$ \\
           & & $^{10}\log$ of ionization parameter & $\log\ [\xi\ (\rm{erg}\ \rm{cm}\ \rm{s}^{-1})]$ & $1.61^{+0.10}_{-0.09}$ \\
           & & Turbulent velocity & $\sigma_v$ & $104^{+34}_{-29}$ \\
           & & Outflowing velocity & $v_{\rm out}\ (\rm{km}\ \rm{s}^{-1})$ & $-240^{+128}_{-100}$ \\
   \hline\xrowht[()]{5pt} 
   Unknown absorber & \texttt{pion} & Hydrogen column density & $N_{\rm H}$ ($10^{20}$ cm$^{-2}$) & $1.74^{+0.64}_{-0.59}$ \\
           & & $^{10}\log$ of ionization parameter & $\log\ [\xi\ (\rm{erg}\ \rm{cm}\ \rm{s}^{-1})]$ & $-0.88\pm 0.19$ \\
           & & Turbulent velocity & $\sigma_v$ & $1114^{+2083}_{-590}$ \\
           & & Outflowing velocity & $v_{\rm out}\ (\rm{km}\ \rm{s}^{-1})$ & $438^{+770}_{-1100}$ \\
   \hline\hline\xrowht[()]{5pt}
   Statistical results & & Best-fit Cash statistic & $C$-stat & 2066.4 \\
    & & Expected Cash statistic & $C$-expt & 1755.9 \\
    & & Degree of freedom & DoF & 1687\\
    \hline
   \end{tabular}
   \end{table*}

\setcounter{table}{0}
\renewcommand{\thetable}{B\arabic{table}}

\setcounter{figure}{0}
\renewcommand{\thefigure}{B\arabic{figure}}

\section{Parameter contour maps of warm absorbers and velocity profiles of absorption lines}
\label{sec:voutprofile}
\begin{figure*}[h!]
\center
\includegraphics[width=\linewidth, clip]{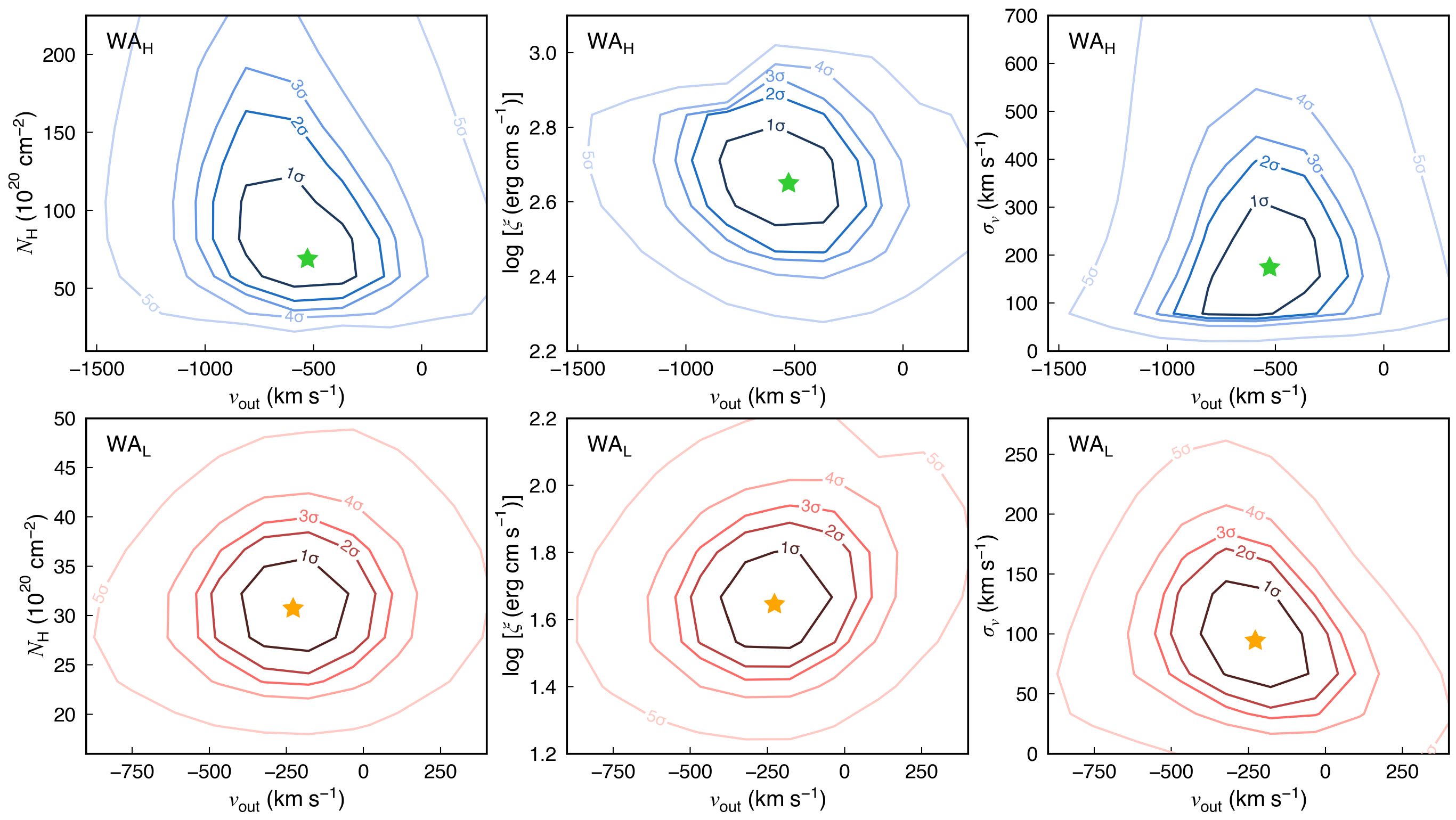}
\caption{
Contour maps of WA$_{\rm H}$ (upper panels) and WA$_{\rm L}$ (lower panels).
The first column indicates outflowing velocity ($v_{\rm out}$) vs. hydrogen column density ($N_{\rm H}$).
The second column indicates $v_{\rm out}$ vs. ionization parameter ($\xi$).
The third column indicates $v_{\rm out}$ vs. turbulent velocity ($\sigma_v$).
Green stars and orange stars show the best-fit parameters for WA$_{\rm H}$ and WA$_{\rm L}$, respectively.
The solid lines in gradient colors from dark to light represent 68.3\% (1$\sigma$), 
90\% (2$\sigma$), 95.4\% (3$\sigma$), 
99\% (4$\sigma$), and 99.99\% (5$\sigma$) confidence levels, respectively.
\label{fig:contourmaps}}
\end{figure*}

\begin{figure*}[h!]
\center
\includegraphics[width=\linewidth, clip]{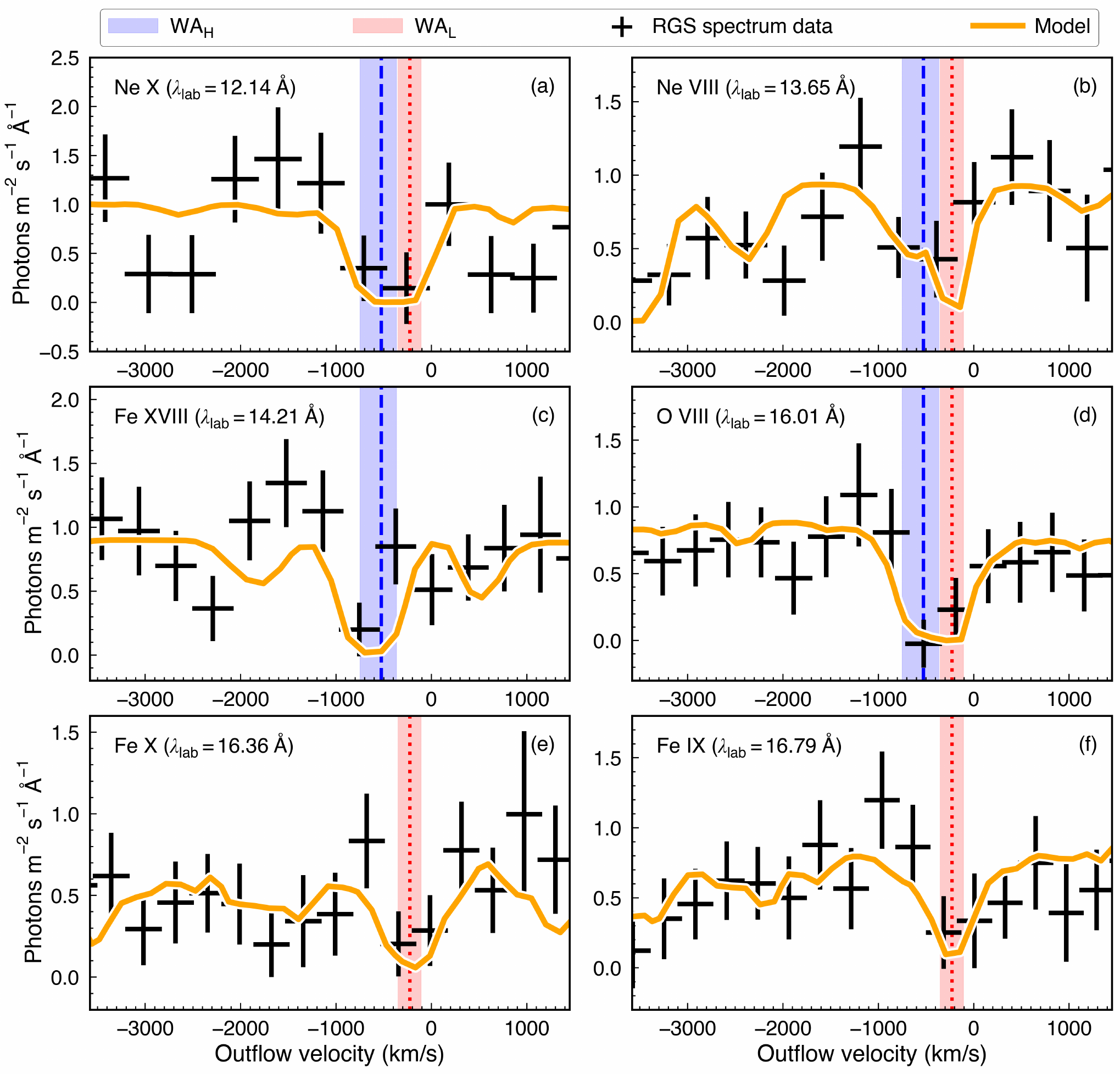}
\caption{
Velocity profiles for Ne X (panel a), Ne VIII (panel b), 
Fe XVIII (panel c), O VIII (panel d), Fe X (panel e), Fe IX (panel f) absorption lines.
The $\lambda_{\rm lab}$ means laboratory wavelength.
The black symbols represent the observed RGS spectrum data with 1$\sigma$ uncertainties.
The solid orange lines denote the best-fit model with \texttt{spex}.
The blue and red lines correspond to the outflowing velocities of WA$_{\rm H}$ and WA$_{\rm L}$, repectively,
while blue and red regions imply their respective 1$\sigma$ uncertainties.
\label{fig:velocityprofile}}
\end{figure*}

Fig. \ref{fig:contourmaps} shows the parameter contour maps of warm absorbers with
confidence level from 68.3\% (1$\sigma$) to 99.99\% (5$\sigma$).
Fig. \ref{fig:RGSspectradetails} illustrates
main absorption lines produced by the two warm absorbers on the RGS spectrum,
while in Fig. \ref{fig:velocityprofile}, we show the velocity profiles for six of these absorption lines.
At the outflowing velocities of the two warm absorbers,
the velocity profiles exhibit significant absorption features.

\setcounter{table}{0}
\renewcommand{\thetable}{D\arabic{table}}

\setcounter{figure}{0}
\renewcommand{\thefigure}{D\arabic{figure}}

\section{Optical spectral fitting with LAMOST spectrum}
\label{sec:LAMOSTspectralfitting}
\begin{figure*}[h!]
\center
\includegraphics[width=\linewidth, clip]{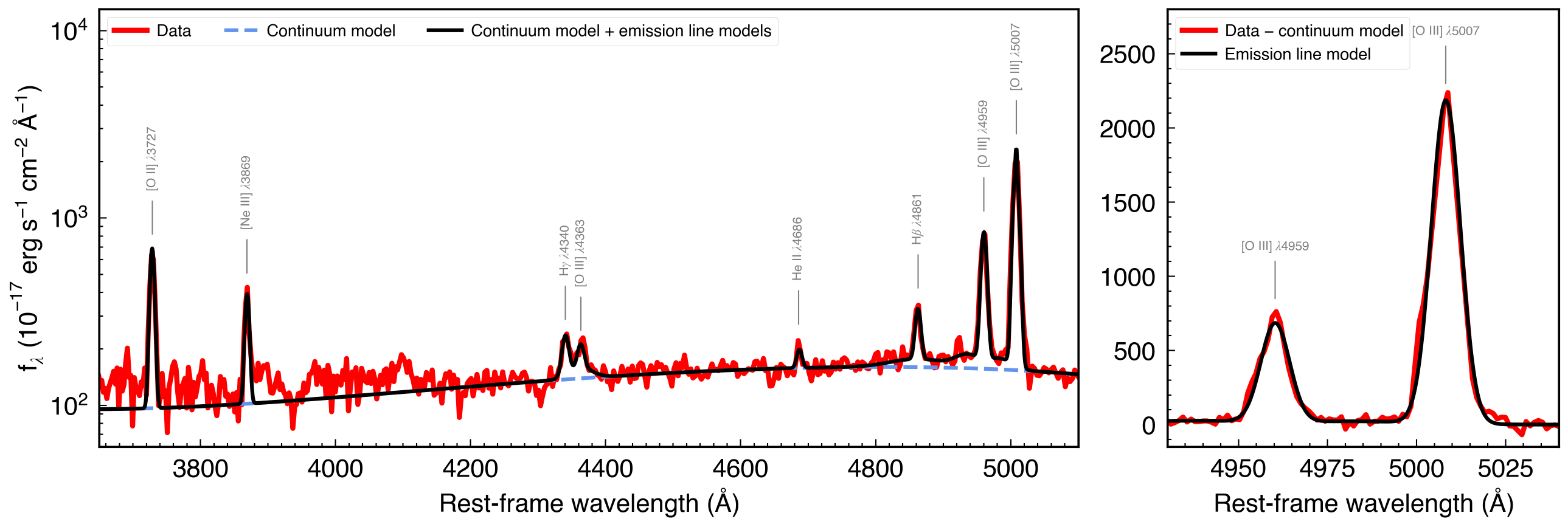}
\caption{LAMOST optical spectrum of 3C~59 and spectral fitting results with PyQSOFit code. Left panel: The solid red line represents the LAMOST spectrum data. The dashed blue line denotes the best-fit continuum model including power-law model and polynomial model accounting for the dust reddening.
The solid black line means the continuum model plus emission line models that are labeled by the gray lines. Right panel: The solid red line shows the spectrum data minus the continuum model. The solid black line represents the emission line models.
\label{fig:LAMOSTspectrum}}
\end{figure*}

The detailed spectral fitting for the LAMOST spectrum of 3C~59 was performed with PyQSOFit code \citep{ShenY2019,GuoHX2018} (see Fig. \ref{fig:LAMOSTspectrum}).
Given that we are only concerned with the [OIII] $\lambda$5007 \AA\ emission line luminosity in this work,
so we just used a simple continuum model that includes a power-law model and a polynomial model accounting for the dust reddening.
The broad and narrow components of emission lines were modeled by multiple Gaussian models.
Each LAMOST spectrum covers the NIR-optical range from 3700 $\AA$ to 9000 $\AA$,
which consists of a blue (3700--5900 $\AA$) and 
a red spectrograph arms (5700--9000 $\AA$)\footnote{See the LAMOST website: \url{http://www.lamost.org/public/node/119?locale=en}}.
In order to estimate the [OIII] $\lambda$5007 \AA\ line luminosity for 3C~59,
we only utilize the blue-arm spectrum data in this work.
The original LAMOST spectrum was only performed with relative flux calibration rather than absolute flux calibration.
Therefore, in this work, we utilized the re-calibrated LAMOST spectrum from Paper I, which is calibrated through matching with SDSS photometric data (see details in Paper I).
In addition, the LAMOST spectrum of 3C~59 had been corrected with the Galactic extinction and the intrinsic extinction from the host galaxy (see detailed introduction about the extinction corrections in Section \ref{sec:NIRoptUVdata}).

\end{appendix}

\end{CJK*}
\end{document}